\documentclass[
 reprint,
 amsmath,amssymb,
 aps,
 superscriptaddress
]{revtex4-2}

\usepackage{filecontents}
\usepackage{braket}
\usepackage{booktabs}
\usepackage{graphicx}
\usepackage{dcolumn}
\usepackage{bm}
\usepackage{xcolor}
\usepackage{svg}
\usepackage[caption=false]{subfig}
\usepackage[export]{adjustbox}
\usepackage{comment}
\usepackage{orcidlink}
\usepackage{xr}

\usepackage{glossaries}
\newacronym{bvk}{BvK}{Born-von-Karman} % chktex 8
\newacronym{hf}{HF}{Hartree--Fock} % chktex 8
\newacronym{tdl}{TDL}{thermodynamic limit} % chktex 8
\usepackage[]{hyperref}

\newcommand{\subfigimg}[3][,]{%
  \setbox1=\hbox{\includegraphics[#1]{#3}}% Store image in box
  \leavevmode\rlap{\usebox1}% Print image
  \rlap{\hspace*{10pt}\raisebox{\dimexpr\ht1-1.5\baselineskip}{#2}}% Print label
  \phantom{\usebox1}% Insert appropriate spcing
}

\begin{document}

\preprint{APS/123-QED}

\title{Exploring the accuracy of the equation-of-motion coupled-cluster band gap of solids}

\def\vienna{%
  \affiliation{%
    Institute for Theoretical Physics, TU Wien,%\\
    Wiedner Hauptstra{\ss}e 8--10/136, 1040 Vienna, Austria % chktex 8
  }%
}

\def\viennahaupt{%
  \affiliation{%
    University of Vienna, Faculty of Physics, %\\
    Kolingasse 14, A-1090 Vienna, Austria
  }%
}

\def\vaspgmbh{%
  \affiliation{%
    VASP Software GmbH, %\\
    Berggasse 21, A-1090, Vienna, Austria
  }%
}

\def\nomadaff{%
  \affiliation{%
    The NOMAD Laboratory at the FHI of the Max-Planck-Gesellschaft%
  }
}

\author{Evgeny Moerman\,\orcidlink{0000-0002-9725-612X}}
\email{moerman@fhi-berlin.mpg.de}
\nomadaff
\author{Henrique Miranda\orcidlink{0000-0002-2843-0876}}
\vaspgmbh
\author{Alejandro Gallo\,\orcidlink{0000-0002-3744-4161}}
\vienna
\author{Andreas Irmler\,\orcidlink{0000-0003-0525-7772}}
\vienna
\author{Tobias Sch\"{a}fer\,\orcidlink{0000-0003-0716-6516}}
\vienna
\author{Felix Hummel\,\orcidlink{0000-0003-1941-3385}}
\vienna
%todo: add Georg Kresse, Henrique Mirandez and Manuel Engel
\author{Manuel Engel\orcidlink{0000-0002-4636-2299}}
\vaspgmbh
\author{Georg Kresse\orcidlink{0000-0001-9102-4259}}
\vaspgmbh
\viennahaupt
\author{Matthias Scheffler\orcidlink{0000-0002-1280-9873}}
\nomadaff
\author{Andreas Gr\"{u}neis\,\orcidlink{0000-0002-4984-7785}}
\vienna

\date{\today}

\begin{abstract}
While the periodic equation-of-motion coupled-cluster (EOM-CC) method promises
systematic improvement of electronic band gap calculations in solids, its practical
application at the singles and doubles level (EOM-CCSD) is hindered by severe finite-size
errors in feasible simulation cells. We present a hybrid approach combining EOM-CCSD with
the computationally efficient $GW$ approximation to estimate thermodynamic limit band gaps
for several insulators and semiconductors.
Our method substantially reduces required cell sizes while maintaining accuracy.
%Our approach achieves unprecendented accuracy.
Comparisons with experimental gaps and self-consistent $GW$
calculations reveal that deviations in EOM-CCSD predictions correlate with reduced
single excitation character of the excited many-electron states.
Our work not only provides a computationally tractable approach to EOM-CC calculations in solids but also 
reveals fundamental insights into the role of single excitations in electronic-structure theory.
%Our work significantly expands the scope of highly accurate methods for the calculation of band gaps in
%solids.
\end{abstract}

\maketitle

%\section{\label{sec:intro}Introduction}
\emph{Introduction.} ---
%%%SHORTENED VERSION
The electronic band gap is an important quantity in semiconductor physics and materials science, \textit{i.e.}
a fundamental measure of a material's optical and electronic properties. 
%It is especially important for renewable energy applications like photovoltaic and thermoelectric devices. 
While density functional theory (DFT) is the state-of-the-art computational method for the electronic
structure of materials, 
its predictive power for band gaps is limited. Local and semi-local
exchange-corelation density functional approximations (DFAs) 
significantly underestimate band gaps~\cite{perdew2017understanding}. Non-local functionals
that include a fraction of exact exchange 
can improve accuracy but often require the tuning of parameters. 
The $GW$ approximation~\cite{hedin1965new,golze2019gw}, particularly in its non-iterative $G_0W_0$ form, 
has emerged as the state-of-the-art method for band structure calculations. 
However, $G_0W_0$ results are sensitive to the choice of the underlying 
DFA, known as the starting-point dependence~\cite{rinke2005combining,fuchs2007quasiparticle}.

To obtain accurate electronic band gaps, more advanced methods that 
systematically incorporate higher orders of electronic correlation effects are needed. Currently, 
one is left with a choice between three methods. The first one consists in the 
self-consistent solution of Hedin's equations ($GW$) in combination with an additional
vertex correction in the screened interaction, that is neglected in $G_0W_0$.
The resulting method, known
as $GW^{\rm TC-TC}$, has been shown to yield good agreement with experimental 
band gaps~\cite{shishkin2007accurate}. One must emphasize that in the $GW^{\rm TC-TC}$ method only the
quasi-particle energies are treated self-consistently while for the orbitals the single-particle
states of the initial mean-field method (in this work the HSE06 DFA~\cite{krukau2006influence}) 
are not updated.
%A self-consistent \emph{GW} treatment has been formulated by 
%Schilfgaarde \emph{et al.}~\cite{van2006quasiparticle} but a vertex-corrected variant is not 
%implemented in the presently used software.
%and to virtually eliminate the starting-point issue 
%of the $G_0W_0$ ansatz.
The high accuracy of $GW^{\rm TC-TC}$ can only be infered by comparing
to experimental gaps properly corrected for zero-point renormalization (ZPR)~\cite{miglio2020predominance, engel2022zero, zacharias2020fully}.
Quantum Monte Carlo (QMC) methods offer an alternative avenue to obtain a highly accurate 
description of electronic correlation. While QMC ground state energies often serve as benchmark results, 
their application to the electronic band gap of materials has been explored
only very recently~\cite{hunt2018quantum, hunt2020diffusion}. Similiarly new to the
field of materials science, is the equation-of-motion coupled-cluster (EOM-CC) method~\cite{stanton1993equation}, which
has been used to study both neutral~\cite{wang2020excitons, lewis2020ab, Gallo2021} and charged
excitations~\cite{mcclain2017gaussian, vo2024performance} in solids. 
In both, the QMC and EOM-CC studies the most challenging obstacle to obtain well converged
band gaps was identified to be the finite-size error.
This error decays relatively slowly with respect to the employed super cell size
or $k$-point mesh used to sample the Brillouin zone
of the primitive unit cell.
To better understand the corresponding scaling law, 
we previously derived the leading order contributions to the finite-size error of 
the EOM-CC band gap~\cite{moerman2024finite}. 
On top of that, we verified numerically that the finite-size error of both the
$G_0W_0$ and the EOM-CCSD band gap converge with the same rate, which motivates
a hybrid approach, in which the convergence behavior can firstly be 
determined using the substantially cheaper $G_0W_0$ method and subsequently be employed for
the extrapolation of the EOM-CCSD band gap. 
Here, we apply this new technique to estimate the fundamental EOM-CCSD band gap of
a number of insulators and semiconductors.
%To understand remaining errors of the EOM-CCSD band gaps, we explore the single excitation
%character of the quasi-particle energies.
% and find a correlation between a 
%decrease of the single excitation character and a notable overestimation of the band gap
%by EOM-CCSD theory. 

%In particular for LiH, we find strong evidence that the
%true theoretical electronic band gap lies actually about $1\,\text{eV}$ above
%the established experimental literature value of $4.9\,\text{eV}$. This finding
%is corroborated by additional vertex-corrected $scGW\Gamma$ calculations and 
%by performing the EOM-CCSD calculation with different approximations to the long-range 
%Coulomb potential with different electronic structure codes.

%\section{\label{sec:theory}Theory}
\emph{Method.} ---
To calculate the electronic band gap in the CC theoretical framework,
we employ the ionization potential (IP)- and electron affinity (EA)-EOM-CC method.
In the case of periodic solids, the IP and EA correspond to the quasi-particle energies
of valence and conduction bands, respectively.
Here, we provide a
succinct summary of the EOM-CC theory. For a detailed account of the
electronic structure factor in EOM-CC theory, which is crucial for the derivation
of the scaling law of the band gap with respect to system size, see
Ref.~\cite{moerman2024finite} and references therein.

%In the EOM-CC methodology, the electronic excitation energies are determined
%as the eigenvalues of
%A central starting point for

The EOM-CC methodology is based on the \textit{similarity-transformed Hamiltonian}
$\bar{H} = e^{-\hat{T}} \hat{H} e^{\hat{T}}$, where $\hat{T}$ is the
\textit{cluster operator} of ground-state CC theory and $\hat{H}$ is
the electronic Hamiltonian.
The wave function of the $n$-th excited state
of the charged system $|\Psi_n^{N\mp1}\rangle$
is obtained by applying an excitation operator $\hat{R}_n^{\text{IP/EA}}$
to $|\Psi_0\rangle$, which is the ground-state wave function of CC theory~\cite{stanton1993equation}:
\begin{eqnarray}\label{eq:linear-ansatz-eom}
 |\Psi_{n}^{N\mp1}\rangle = \hat{R}_n^{\text{IP/EA}}|\Psi_{0}\rangle\text{.}
\end{eqnarray}
The $\hat{R}_n^{\text{IP}}$ and $\hat{R}_n^{\text{EA}}$ operators in the EOM-CC method with single-
and double excitations (EOM-CCSD) are
%\begin{subequations}
%    \begin{eqnarray}\label{eq:ip-operator}
$        \hat{R}^{\text{IP}}_{n} =
        \sum_{i} r_{i,n}\hat{a}_{i} +
        \sum_{ija} r^{a}_{ij,n}\hat{a}^{\dag}_{a}\hat{a}_{i}\hat{a}_{j} $
and
%    \end{eqnarray}
%    \begin{eqnarray}\label{eq:ea-operator}
$        \hat{R}^{\text{EA}}_{n} =
        \sum_{a} r^{a}_{n}\hat{a}^{\dag}_{a} +
        \sum_{iab} r^{ab}_{i,n}\hat{a}^{\dag}_{a}\hat{a}^{\dag}_{b}\hat{a}_{i}$,
%    \end{eqnarray}
%\end{subequations}
respectively.
The indices $i$, $j$ and $a$, $b$ denote occupied and virtual
orbitals. In the case of Bloch orbitals, the orbital index serves as compound index
for the $k$-vector of the primitive unit cell and band number.

IP/EA-EOM-CCSD excitation energies are determined as the eigenvalues of $\bar{H}$:
\begin{eqnarray}
        \bar{H}(\hat{R}_{n}^{\text{IP/EA}}|\Phi_0\rangle) = E_n^{\text{IP/EA}}(\hat{R}_{n}^{\text{IP/EA}}|\Phi_0\rangle)\text{,}
\end{eqnarray}
whereas $r_{i,n}$, $r^{a}_{ij,n}$  and $r^{a}_{n}$, $r^{ab}_{i,n}$
represent excited many-electron states.
The fundamental electronic band gap is given by the sum 
between the first IP and EA excitation energies 
\begin{eqnarray}
	E_{\text{gap}}^{\text{EOM}} = E^{\text{IP}} + E^{\text{EA}}\text{.} 
\end{eqnarray}
Note that we have dropped the state index $n$ for brevity.
One useful metric to classify EOM-CC excitations is the \textit{single excitation character} defined as

\begin{eqnarray}\label{eq:def-r1}
	n_1^{\text{IP}} = \sum_{i} r_{i}^2\quad
	n_1^{\text{EA}} = \sum_{a} (r^{a})^2\quad\text{.}
\end{eqnarray}
Especially for low-order truncations of EOM-CC theory, like the EOM-CCSD method,
values of $n_1^{\text{IP/EA}}$ close to $1$ are 
associated with an increased accuracy~\cite{ranasinghe2019vertical}.
%Ranasinghe \emph{et al.}~\cite{ranasinghe2019vertical} discussed a decrease in single excitation character 
%as a marker for an increase in orbital relaxation effects, which are challenging to capture for 
%low-order EOM-CC methods.

% todo: add gap definitio E_{\text{gap}}=... --DONE

%the $n$-th IP and EA excitation energies correspond to quasi-particle energies of the $n$-th valence 
%and conduction band and their difference to the electronic band gap.

In this work we employ an efficient extrapolation approach that links
the scaling of EOM-CC and $G_0W_0$ gaps with respect to super cell size.
We first perform a series of
$G_0W_0$ band gap calculations sampling the first Brillouin zone employing a $k$-mesh with a total of
$N_k$ points.
These $G_0W_0$ band gaps are fitted to
%expression~\cite{moerman2024finite} for three-dimensional systems given by
\begin{eqnarray}\label{eq:3d-convergence-rate}
	E_{\text{gap},N_k}^{G_0W_0} =
	E_{\text{gap,TDL}}^{G_0W_0} + 
	A N_k^{-\frac{1}{3}} + B N_k^{-\frac{2}{3}} + C N_k^{-1}.
	\text{.}
\end{eqnarray}
Although the above equation was derived for EOM-CCSD band gaps ($E_{\text{gap}}^{\text{EOM}}$),
we find that it works well for $G_0W_0$~\cite{moerman2024finite}.
This can partly be attributed to the fact that the $G_0W_0$ and EOM-CCSD methods are closely related,
which was elucidated in Refs.~\cite{lange2018relation,tolle2023exact}.
We emphasize that there exist well established corrections to the finite-size error in the $G_0W_0$ band gap,
which are referred to as head- and wing-corrections~\cite{hybertsen1987ab}.
Here, however, we take advantage of the similar finite-size scaling in $G_0W_0$ and EOM-CCSD theory by proceeding without these corrections.
In the thermodynamic limit (TDL), that is for $N_k\to\infty$, the $G_0W_0$ and EOM-CCSD band gaps
are strictly linked by~\cite{moerman2024finite}
\begin{eqnarray}\label{eq:eom-vs-gw-linear-relation}
E_{\text{gap},N_k}^{\text{EOM}} = a + b\cdot E_{\text{gap}, N_k}^{G_0W_0}\text{.}
\end{eqnarray}
The assumption of the present approach is that
the above equation already holds approximately for smaller system sizes, which
is corroborated by the findings summarized in the next section.
Once the system specific fit parameters $a$ and $b$ are obtained using relatively
small $N_k$, one can estimate the
EOM-CCSD gap in the TDL by $E_{\text{gap},\text{TDL}}^{\text{EOM}} = a + b\cdot E_{\text{gap},\text{TDL}}^{G_0W_0}$, which
follows from the linear scaling relation given above.
We note that for $G_0W_0$ calculations $k$-summations are performed using $k$-meshes
that correspond to the super cells used for the EOM-CCSD calculations.
Furthermore the $G_0W_0$ calculations in the above procedure employ the same \gls{hf} single-particle
energies and wave functions ($G_0W_0$@HF).
%It will be shown that for the four materials studied in this work, a linear correlation between the
%band gap convergence of both methods can be observed. Knowledge of this linear correlation in combination
%with the previously determined converged $G_0W_0$@HF band gap, provides a straight-forward opportunity
%to determine the EOM-CCSD band gap in the TDL. For all CC, EOM-CC and $GW$ calculations performed

%\section{Results}\label{sec:results}
\emph{Results.} ---
\begin{figure}[t]
    \centering
   \begin{center}
    \includegraphics[width=4.5cm, height=3.5cm]{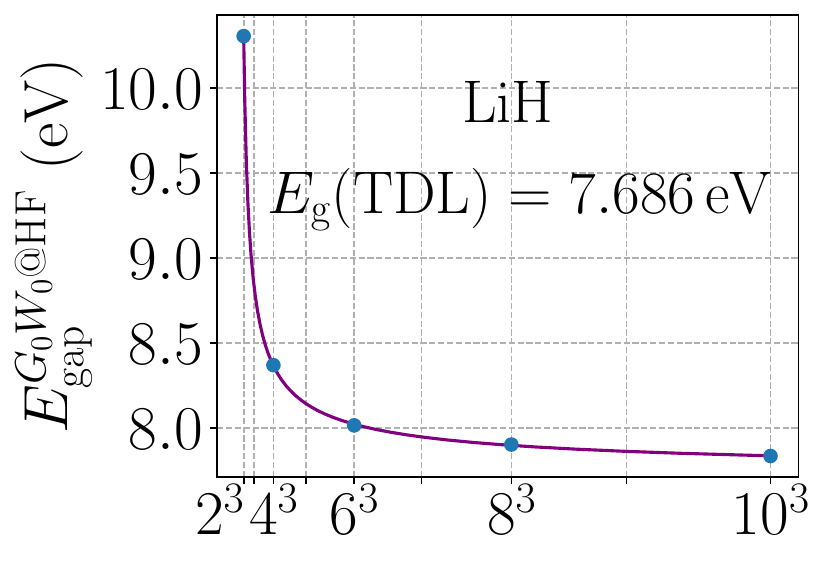}%
	   \raisebox{0.1cm}{\includegraphics[width=4.2cm, height=3.38cm]{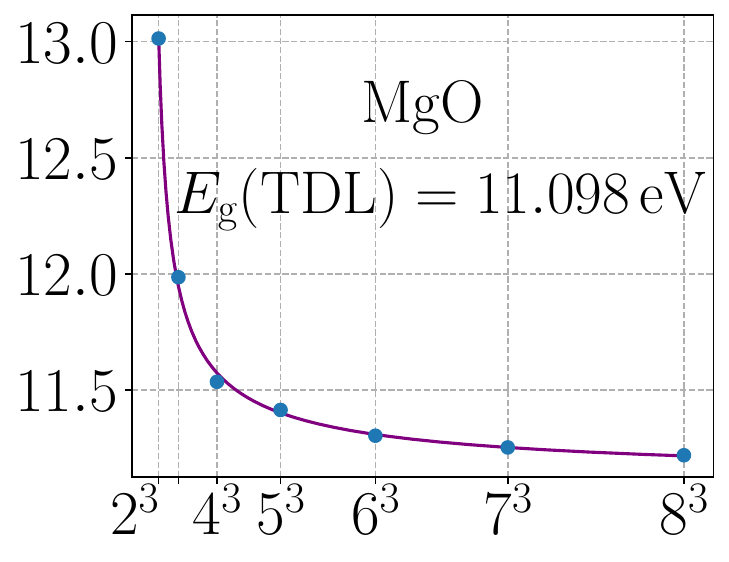}}
    \includegraphics[width=4.5cm, height=3.8cm]{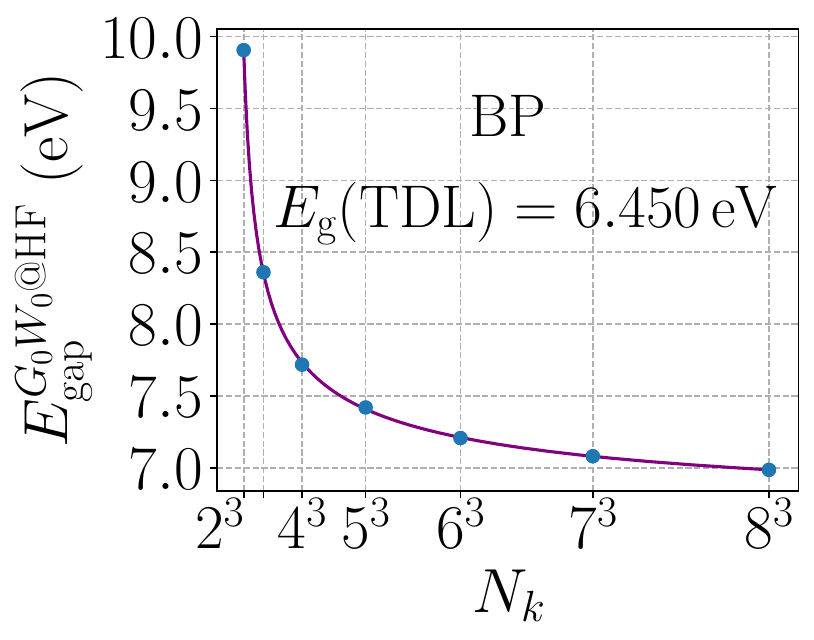}%
    \includegraphics[width=4.2cm, height=3.7cm]{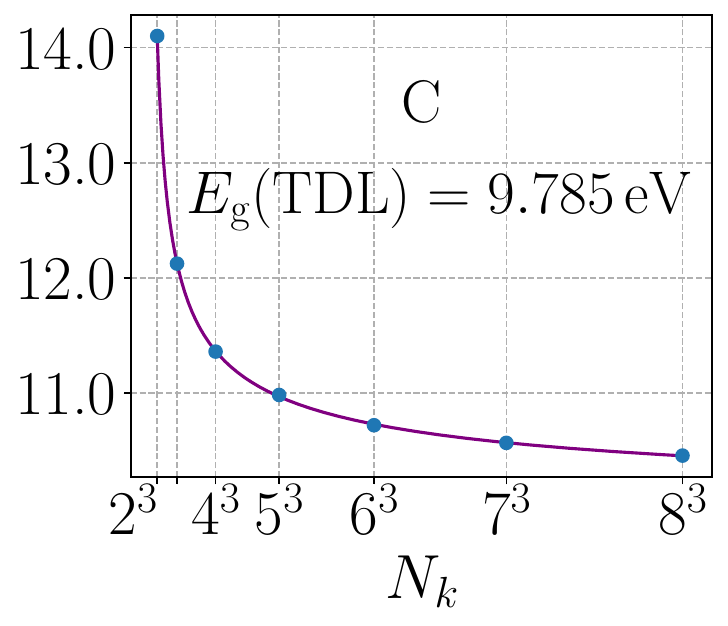}
    \end{center}
    \caption{Scaling of the $G_0W_0$@HF band gaps
	of LiH, MgO, BP and C with respect to the $k$-mesh using FHI-aims (NAOs). The data points were fitted and
    extrapolated using the expression 
	$E_{\mathrm{gap},N_k}^{G_0W_0\text{@HF}} = E_{\mathrm{gap},\mathrm{TDL}}^{G_0W_0\text{@HF}} + AN_k^{-1/3} + BN_k^{-2/3} + CN_k^{-1}$ (see Eq.~(\ref{eq:3d-convergence-rate})). $E_{g}(\text{TDL})$ is a shorthand for $E_{\mathrm{gap},\mathrm{TDL}}^{G_0W_0\text{@HF}}$, \textit{i.e.} using a converged $k$ summation.}
    \label{fig:3d-gw-extrapolations}
\end{figure}
We now assess the system size convergence of the $G_0W_0$ band gaps and their extrapolation to the TDL
based on Eq.~(\ref{eq:3d-convergence-rate}).
%In the first step of the herein presented methodology, we want to determine the band gap of the $G_0W_0$ method in the TDL
%via extrapolation.
Figure~\ref{fig:3d-gw-extrapolations} shows that the computed $G_0W_0$ band gaps are well approximated
by Eq.~(\ref{eq:3d-convergence-rate}) previously derived for IP/EA-EOM-CCSD energies, indicating
that both methods yield band gaps that converge to the TDL with the same scaling behaviour.
This is supported by Lange and Berkelbach\cite{lange2018relation}, who showed that
the $G_0W_0$@HF and  IP- and EA-EOM-CCSD approximations feature identical low-order ring terms
that also play an important role for long-range correlation effects in the ground state~\cite{mattuck}.
In contrast to
the ground-state CC correlation energy, which is known to converge to the TDL with a $1/N_k$ rate for 
insulators~\cite{gruber2018applying,xing2024inverse}, our
analsysis suggests in general a $1/N_k^{1/3}$ leading-order behavior of the band gap in the large $N_k$ limit~\cite{moerman2024finite}.
Moreover, additional next-to-leading-order contributions  need to be included to model the 
band gap convergence for relatively small $k$-meshes.
%We stress again that $G_0W_0$@HF band gaps including head- and wing-corrections converge faster to the TDL.
%However, since we obtain our estimate of the finite-size converged $G_0W_0$@HF band gap via extrapolation,
%we don't make use of such corrections.
The results above have been obtained using FHI-aims 
(see the SI for computational details).

\begin{figure}[t]
    \centering
   \begin{center}
    \includegraphics[width=4.5cm, height=3.5cm]{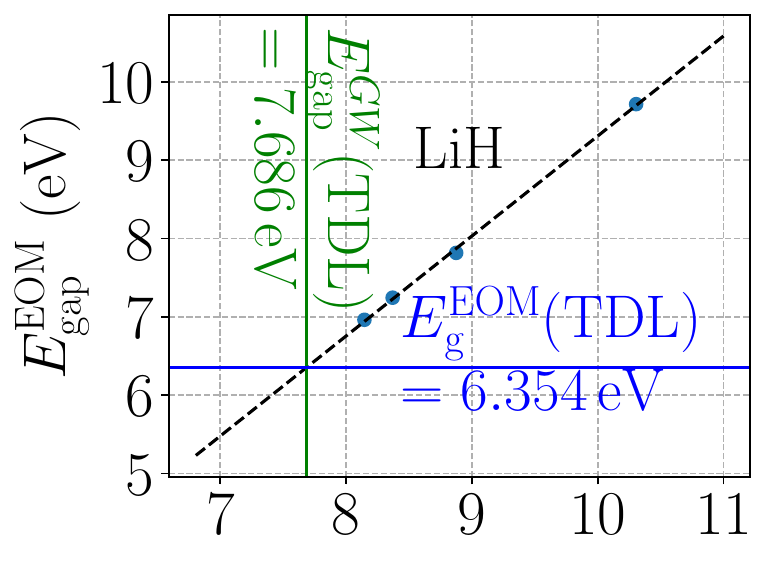}%
    \includegraphics[width=4.2cm, height=3.5cm]{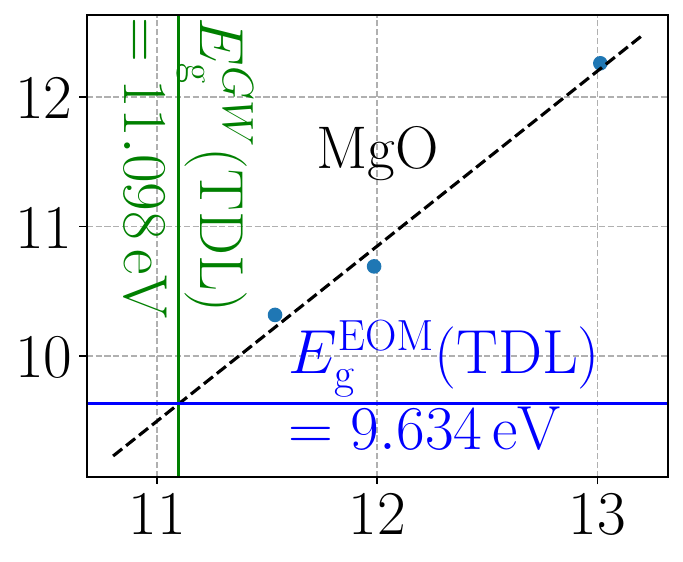}
    \includegraphics[width=4.5cm, height=4.0cm]{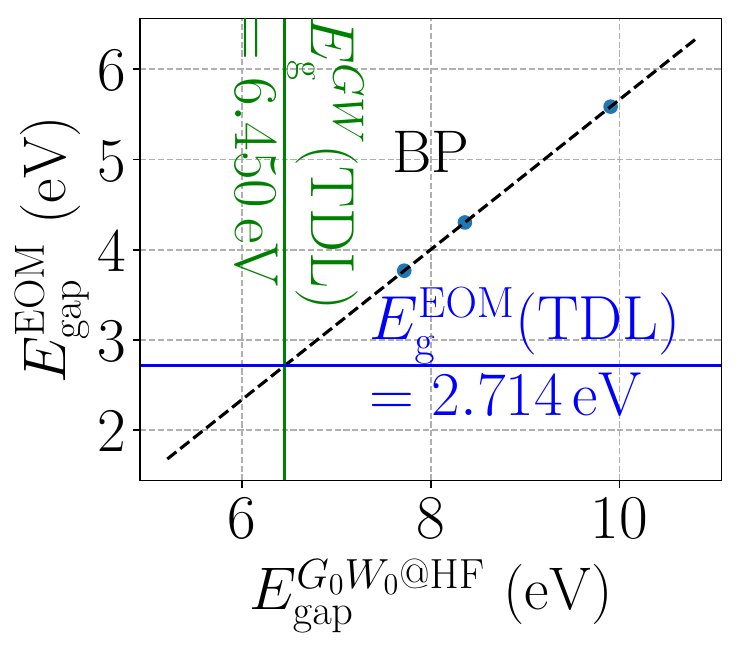}%
    \includegraphics[width=4.2cm, height=4.0cm]{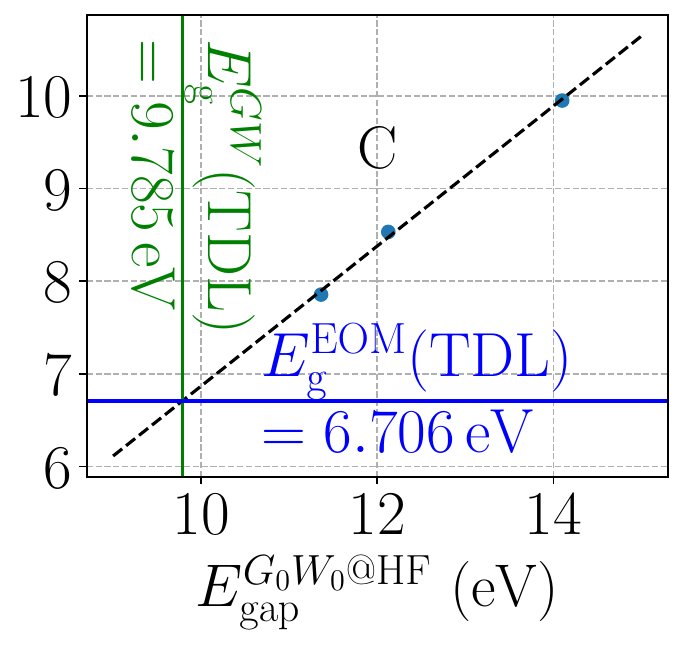}
    \end{center}
    \caption{Correlation of the EOM-CCSD and $G_0W_0$@HF band gap
    convergence of LiH, MgO, BP and C
	with increasing super cell size (EOM-CCSD) and $k$-mesh density ($G_0W_0$@HF) using FHI-aims. 
    %For LiH
    %isotropic supercells of size $2\times2\times2-5\times5\times5$ were used, while for the
    %other materials $2\times2\times2-4\times4\times4$ supercells were employed. 
    %For all four
    %systems the band gap converges from above, so that system size increases
    %from right to left.
    The extrapolated TDL value of the $G_0W_0$@HF and of the EOM-CCSD band gap is denoted by $E_{\text{g}}^{\text{GW}}(\text{TDL})$
    and $E_{\text{g}}^{\text{EOM}}(\text{TDL})$ and marked by green and blue lines, respectively.
    The number of $k$-points increases from $2\times2\times2$, over $3\times3\times3$ to $4\times4\times4$.
    For LiH results for $5\times5\times5$ are also shown.}
    \label{fig:3d-eom-gw-correlation}
\end{figure}

The next step in the procedure outlined above lies in 
determining how the scaling behaviour of the $G_0W_0$@HF band gap and
the EOM-CCSD band gap are correlated with each other.
Figure \ref{fig:3d-eom-gw-correlation} shows the band gaps of the two methods
%($2\times2\times2 - 4\times4\times4$ supercells for MgO, BP and C and $2\times2\times2 - 5\times5\times5$ for LiH)
plotted against each other and fitted according to Eq.~(\ref{eq:eom-vs-gw-linear-relation}).
Indeed, we find that the data points of the four materials consistently lie on a straight line.
The validity of the herein established relation between the two methods is most clear in the case of LiH, where
a fourth data point ($5\times 5\times 5$) is accessible due to the smaller number of electrons per unit cell. For LiH, 
all four points align on a straight line. 
This enables a direct estimation of the EOM-CCSD band gap in the TDL given by
$E_{\text{gap},\text{TDL}}^{\text{EOM}} = a + b\cdot E_{\text{gap},\text{TDL}}^{G_0W_0}$
and depicted in the plots of Figure \ref{fig:3d-eom-gw-correlation}
at the intersection point between the linear fit and the green vertical line, which shows $E_{\text{gap},\text{TDL}}^{G_0W_0\text{@HF}}$ determined previously in Figure \ref{fig:3d-gw-extrapolations}.
Let us emphasize that our approach only depends on the scaling behavior of the $G_0W_0$@HF gap,
its absolute value is irrelevant.
For C, BP and MgO we find that the extrapolated EOM-CCSD band gap value changes at most
$0.3\,\text{eV}$ if the extrapolation is performed using the $2\times2\times2$ and $3\times3\times3$ data points only, omitting the biggest super cell size of $4\times4\times4$. 
To confirm the correctness of the IP- and EA-EOM-CCSD implementation in combination with FHI-aims, a small molecular benchmark was conducted and compared to published results (see Table \ref{supp-tab:molecules-ip} and \ref{supp-tab:molecules-ea}).

\begin{table}
	\centering
	\caption{
		Comparison of EOM-CCSD band gaps calculated using different computer codes: 
		NAO-based FHI-aims and PAW-based VASP results obtained from the linear fit against $GW$ gaps as shown in 
		Figure \ref{fig:3d-eom-gw-correlation} and \ref{supp-fig:3d-eom-gw-correlation-vasp} 
		and corrected for basis set incompleteness (see discussion in Section \ref{supp-sec:bsie}).
		PySCF results were taken from Reference~\cite{vo2024performance}. 
		$GW^{\rm TC-TC}$ results and experimental 
		values
		%~\cite{baroni1985quasiparticle,whited1973exciton, chiang1989electronic, peter2010fundamentals} 
		are shown for reference. 
		%For experimental gaps 
		%obtained from photoelectron spectroscopy, 
		The zero-point renormalized
		value (w/o ZPR) is shown next to the experimentally observed (obs.) one.
		All values in eV.
		%Summary of the
		%EOM-CCSD band gaps.
		%The EOM-CCSD band gap results based on NAOs (FHI-aims) and the PAW method (VASP) have been obtained
		%from the linear fit against $GW$ results as shown in Figure \ref{fig:3d-eom-gw-correlation} (FHI-aims)
		%and \ref{supp-fig:3d-eom-gw-correlation-vasp} (VASP) and corrected for the basis set incompleteness
		%error tabulated and discussed in Section \ref{supp-sec:bsie}. 
		%Additionally, results based on Gaussian Type Orbitals (GTO) via PySCF are shown for comparison~\cite{vo2024performance}.
		%The fifth column contains the electronic band gaps obtained using the $GW^{\rm TC-TC}$ method.
		%The final column shows experimental band gap values~\cite{baroni1985quasiparticle,whited1973exciton, chiang1989electronic, peter2010fundamentals} with the zero-point renormalization 
		%(ZPR) subtracted (w/o ZPR) next to the experimentally observed (obs.) value.
                %Only photoelectron spectroscopy measurements are corrected for ZPR.
		%All entries are given in eV.
		}
	\label{tab:3d:gap-error-summary}
	\begin{tabular}{llllllc}
		\toprule
		\multicolumn{1}{ c }{\footnotesize Material}&
		\multicolumn{3}{ c }{EOM-CCSD}&
		\multicolumn{1}{ c }{\footnotesize $GW^{\rm {\tiny TC-TC}}$} &\multicolumn{1}{ c }{Exp.}\\
		\cmidrule{2-6}
		    &
		\footnotesize VASP$^{\rm a}$ &
		\footnotesize FHI-aims$^{\rm b}$ &
		\footnotesize PySCF &
		\footnotesize VASP &
		\footnotesize w/o ZPR\,(obs.) \\
		\midrule
		LiH  & 6.25  & 6.32 & 5.85 & 5.52 & 5.43\,(4.99$^{\rm g}$)\\
		C    & 5.75  & 6.15 & 4.88 & 5.88 & 5.80\,(5.48$^{\rm c}$)\\
		BP   & 2.27  & 2.38 & 1.65 & 2.19 & 2.26\,(2.16$^{\rm f}$)\\
		MgO  & 9.52  & 9.19 & 8.34 & 8.10 & 8.36\,(7.83$^{\rm j}$)\\
		\midrule
		Si   & 1.29  &      & 0.93 & 1.24 & 1.23\,(1.17$^{\rm d}$)\\
		BN   & 6.62  &      & 6.45 & 6.58 & 6.5\,(6.1$^{\rm e}$)\\
		LiF  & 16.19 &      & 15.43& 14.73 & 15.43\,(14.2$^{\rm h}$)\\
		LiCl & 9.90  &      & 9.43 & 9.53 & 9.94\,(9.40$^{\rm i}$)\\
	\bottomrule
	\end{tabular}
	\begin{flushleft}
	$^{\rm a}$ Extrapolating from $n\times n\times n$ super cells with $n$=2 and 3\\
	$^{\rm b}$ Extrapolating from $n\times n\times n$ super cells with $n$=2, 3 and 4\\
	$^{\rm c}$\cite{chiang1989electronic}, $^{\rm d}$\cite{madelung2004semiconductors},
	$^{\rm e}$\cite{levinshtein2001properties}, $^{\rm f}$\cite{woo2016bp},
	$^{\rm g}$\cite{lushchik1977electronic, baroni1985quasiparticle}, 
	$^{\rm h}$\cite{PhysRevB.13.5530},
	$^{\rm i}$\cite{baldini1970optical}, $^{\rm j}$\cite{whited1973exciton}
        
	\end{flushleft}
\end{table}

We now turn to the discussion comparing  EOM-CCSD band gaps obtained with different implementations.
Specifically, we have carried out EOM-CCSD calculations employing electronic Hamiltonians
computed by FHI-aims and VASP.
These are interfaced implementations~\cite{moerman2022interface, hummel2017low} using 
super cells, equivalent to corresponding $k$-meshes of the primitive unit cell.
Additionally, we compare our findings against EOM-CCSD results from a previous PySCF study
by Vo, Wang and Berkelbach~\cite{vo2024performance}.
PySCF calculations used pseudopotentials optimized for Hartree–Fock and Gaussian-type orbitals (GTOs).
VASP employs the frozen core approximation with DFT-PBE core states, but including core-valence exact exchange~\cite{Paier2005}.
The present FHI-aims calculations employ the frozen core approximation only on the level of post-HF theories.
Therefore core-valence correlation is neglected in all  summarized results, whereas
the treatment of core-valence exchange exhibits small inconsistencies.
%The results of the other two codes
%employ approximations for the description of the core electrons.
In the complete basis set and thermodynamic limit, the band gaps for the same systems
should agree to within remaining uncertainties that derive from, \textit{e.g.}, differences in the frozen core approximation.
Indeed, we find that
% $G_0W_0$@HF band gaps extrapolated to the TDL and obtained by
VASP and FHI-aims agree well with each other.
% to within around 0.1-0.3\,eV for LiH, MgO, C and BP (see SI).
%Data on the convergence of the $G_0W_0$@HF and 
%EOM-CCSD band gaps calculted using VASP can be found in the SI.
%Data 
%For a more extensive comparison we have performed additional calculations using the VASP-PAW method with a two-point
%finite-size extrapolation, which was shown to yield reasonable agreement with more rigorously
%converged FHI-aims results. These EOM-CCSD gaps are given in Table~\ref{tab:3d:gap-error-summary}
%and exhibit similarly large discrepancies with GTO-based findings
%as NAO-based EOM-CCSD gaps discussed above.
Table~\ref{tab:3d:gap-error-summary} shows that for LiH and BP our EOM-CCSD band gaps
computed with FHI-aims and VASP yield excellent agreement to within  0.1\,eV.
Only for MgO and C, VASP and FHI-aims exhibit discrepancies of about 0.3-0.4\,eV.
This can be partly attributed to the fact that VASP
calculations employed $2\times2\times2$ and $3\times3\times3$ super cells
only due to computational constraints.
%However, these smaller cells enable calculations for a larger number of systems.

Table~\ref{tab:3d:gap-error-summary} reveals that 
the  differences between the FHI-aims and VASP gaps compared to the PySCF results are significant.
For LiH, C, MgO and BP the average absolute difference of the EOM-CCSD gaps between VASP/FHI-aims and PySCF is 0.8\,eV,
whereas VASP and FHI-aims agree to within 0.2\,eV on average.
What is the reason for these significant discrepancies?
Both works have tried to converge the computed band gaps with respect to accessible basis set and system size.
For example, Fig.~2 of Ref.~\cite{vo2024performance} depicts TDL extrapolations of EOM-CCSD
band gaps for different basis sets. In agreement with our findings, the
basis set convergence of EOM-CCSD band gaps is relatively fast, indicating that the
most likely source of the discrepancy are finite-size errors.
Both TDL extrapolations shown in our Fig.~\ref{fig:3d-eom-gw-correlation} and 
Fig.~2 from Ref.~\cite{vo2024performance} look reliable at first sight for the system sizes studied
($2\times2\times2$ to $4\times4\times4$).
However, we stress that the $N_k^{-1/3}$ TDL extrapolations in Ref.~\cite{vo2024performance} assume
that the studied systems are already large enough such that the band gap convergence is
dominated by the corresponding leading-order finite-size error.
Yet, based on numerical findings for low-dimensional systems, we expect that
system sizes of about $8\times8\times8$ would be needed to observe a convergence that follows
the pure $N_k^{-1/3}$ behavior~\cite{moerman2024finite}.
%Hence extrapolations using $2\times2\times2$ to $4\times4\times4$ are not  reliable.
In fact, employing a $N_k^{-1/3}$ TDL extrapolation of the EOM-CCSD gaps for LiH
with VASP and FHI-aims would also exhibit a large discrepancy on the scale of more than 1\,eV (see SI).
%This is caused by the fact that FHI-aims employs a spherical Coulomb truncation technique,
%while for the VASP calculations we use the probe-charge Ewald approximation.
%Nonetheless, the $G_0W_0$-aided approach presented
%in this work appears to be robust independently of the underlying approximation of the Coulomb potential,
%as EOM-CCSD band gaps for FHI-aims and VASP agree very well.
%%This is confirmed especially in the case of LiH, which w validated using a $5\times5\times5$ cell.
%%as FHI-aims (NAO) employs a spherical truncation technique, while for the VASP calculations we have decided to use 
%%the probe-charge Ewald approximation. 
%%is not achieved using a conventional extrapolation to the TDL as done in Ref.~\cite{vo2024performance}.
We also note that the approach of Ref.~\cite{vo2024performance} yields band gaps that converge
from below with increasing $k$-mesh densities, whereas our band gaps converge from above.
The reason for this different behavior in Refs.~\cite{vo2024performance,mcclain2017gaussian}
is that the underlying HF calculations neglect the integrable singularity contribution of the Coulomb potential
in the exchange operator, leading to underestimated HF band gaps.
We argue that the finite-size errors in the EOM-CCSD band gap of Ref.~\cite{vo2024performance}
are still dominated by the underlying HF finite-size errors, which only partly cancel with EOM-CCSD finite-size errors.
In contrast, our finite-size errors originate from post-HF EOM-CCSD terms, which decrease
the HF band gap for increasing super cell sizes.
The treatment of the singularity of the Coulomb potential
strongly affects the finite-size convergence behavior of HF~\cite{sundararaman2013regularization}
and CC~\cite{mcclain2017gaussian,xing2024inverse} theories.
To summarise, we conclude that our $GW$-scaling EOM-CCSD extrapolation technique
efficiently compensates finite-size errors and that the obtained
band gaps are more precise than those obtained in a previous study by Vo \textit{et al.}~\cite{vo2024performance}.

%We will continue to underpin this statement in the following discussion.

\begin{figure}
	\captionsetup[subfigure]{labelformat=empty}
  \centering
  %\begin{tabular}{@{}p{0.45\linewidth}@{\quad}p{0.45\linewidth}@{}}
	  \subfloat[\label{fig:h2-chain-finite-size-convergence}]{\subfigimg[width=4.5cm, height=3.5cm, valign=t]{a)}{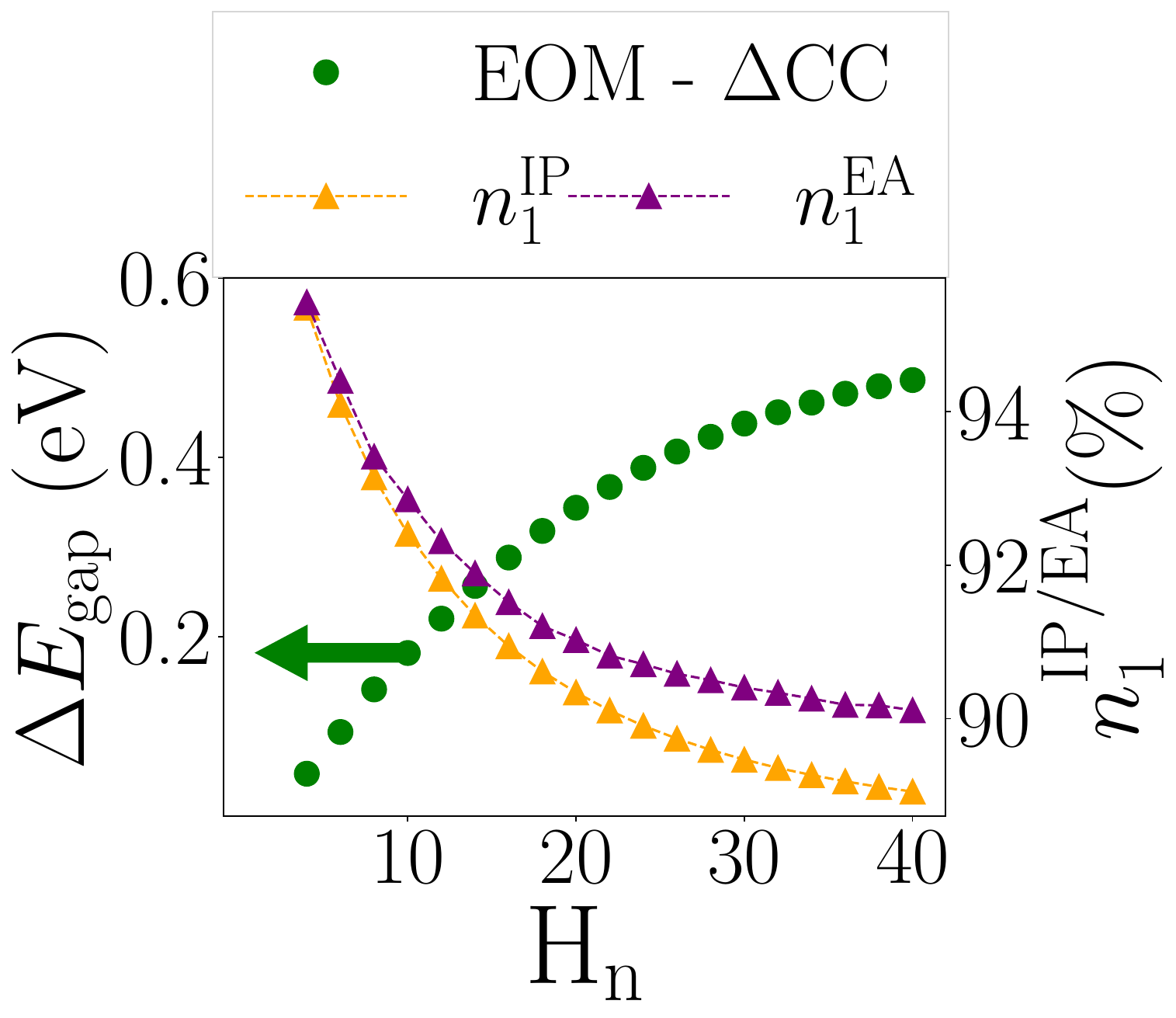}}
	  \subfloat[\label{fig:n1ip-vs-error-3d}]{\subfigimg[width=4.5cm, height=3.5cm, valign=t]{b)}{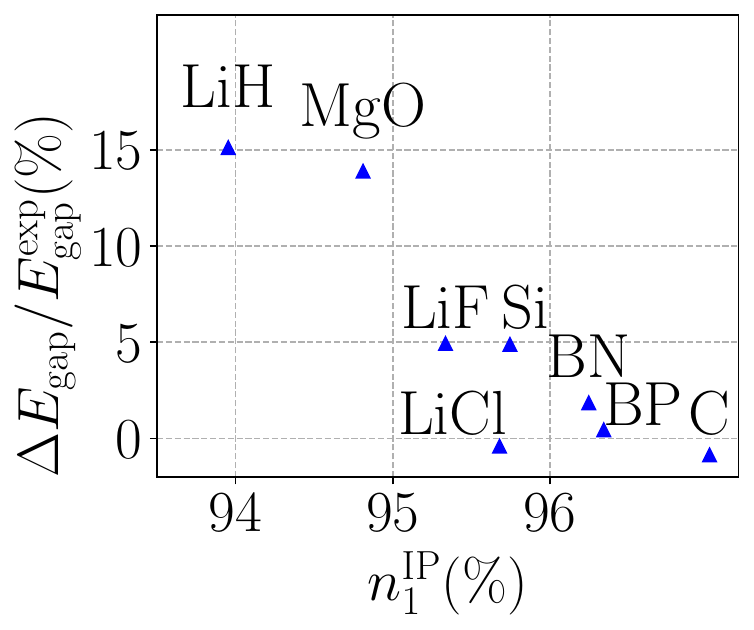}}
    %\subfigimg[width=3.5cm, height=3.5cm, valign=t]{a)\label{fig:r1-vs-nk-ip-3d}}{figures/r1_ip_vs_system_size_3D.pdf} &
    %\subfigimg[width=4.5cm, height=3.5cm, valign=t]{b)\label{fig:h2-chain-finite-size-convergence}}{figures/eom_vs_delta_ccsd_t_h2_chain.pdf}
  %\end{tabular}
	\caption{
	(a) Single excitation character and gap deviation of EOM-CCSD relative to $\Delta$CCSD(T) gap
	for the molecular hydrogen chain with respect to the chain length.
        (b)Relation between relative error of EOM-CCSD gap compared to experiment 
	and single excitation character.
}
  \label{fig:r1_vs_nk_h2_and_3D}
\end{figure}

Before discussing the accuracy of EOM-CCSD band gaps,
we briefly focus on $GW^{\rm TC-TC}$ and the experimental band gaps
summarized in Table~\ref{tab:3d:gap-error-summary}.
$GW^{\rm TC-TC}$ calculations were performed using the projector augmented wave (PAW) method as 
implemented in VASP (see the SI for computational details)
and are in excellent agreement with ZPR corrected experimental band gaps
for all studied systems, confirming Refs.~\cite{miglio2020predominance,engel2022zero}.
The employed ZPR correction is explained in Ref.~\cite{engel2022zero} and computational details can be found in the SI.
In agreement with recent findings, the ZPR corrections are significant for C and MgO~\cite{miglio2020predominance}.
Here, we show that the ZPR corrections for LiH are also on the scale of about 0.4\,eV,
substantially larger than previously reported in Ref.~\cite{monserrat2013anharmonic} and in agreement with experiment~\cite{plekhanov1997isotopic}.
%Note that the gaps of LiH, LiF, LiCl and MgO are not corrected for ZPR because they are inferred from optical
%measurements.
%For completeness we also report the ZPR corrections of these systems in the SI.
The good agreement between $GW^{\rm TC-TC}$ and (ZPR corrected) experiment
supports our confidence in the correctness of the experimental gaps.

We now turn to the discussion of the accuracy of EOM-CCSD gaps compared to experiment.
Note that FHI-aims EOM-CCSD gaps were obtained using larger super cell sizes to 
verify the reliability of the computationally cheaper VASP EOM-CCSD calculations using
smaller super cell sizes. The latter approach could also be applied to a larger number of systems.
It is noteworthy that our EOM-CCSD gaps are in excellent agreement
with ZPR corrected experimental gaps for C, Si, LiCl, BN and BP.
However, the large errors of EOM-CCSD band gaps for LiH, LiF and MgO are
unexpected given the accuracy achieved by IP-EOM-CCSD for small molecules compared
to experiment, where mean absolute deviations of about 0.15\,eV have been observed~\cite{ranasinghe2019vertical}.
Is the EOM-CCSD approximation less accurate for periodic systems than for molecular systems?
To address this question, we investigate the single excitation character
of the EOM-CCSD states, $n_1^{\text{IP/EA}}$, which has been used in 
previous studies to explain the performance of a variety of CC methods for molecular 
IPs~\cite{ranasinghe2019vertical, shaalan2022accurate, marie2024reference, stanton1999simple} and 
neutral double excitations~\cite{kossoski2024reference}.  
%To the best of our knowledge this relation has not yet been investigated for periodic systems. 
%%%SHORTENED VERSION
Before discussing our findings for periodic systems, we demonstrate the significance of $n_1^{\text{IP/EA}}$
using the hypothetical example of dimerized hydrogen chains ($\text{H}_4$ to $\text{H}_{40}$).
We compare band gaps calculated via EOM-CCSD against 
$\Delta$CCSD(T) - a reference method~\cite{lange2018relation,krause2015coupled} that explicitly computes 
the ground-state energy of the charged $n+1$ and $n-1$-electron systems 
thus capturing orbital relaxation effects upon addition/removal of an electron. Using a 20\% 
bond length alternation to ensure insulating behavior, Figure~\ref{fig:h2-chain-finite-size-convergence}
clearly demonstrates that the agreement of EOM-CCSD with $\Delta$CCSD(T) 
deteriorates from $50\,\text{meV}$ for $\text{H}_4$ to $0.5\,\text{eV}$ for $\text{H}_{40}$. 
This deviation correlates with a decrease in $n_1^{\text{IP/EA}}$ from 95\% to about 90\%, 
highlighting the limitations of EOM-CCSD for extended systems.

Finally, we investigate the relationship between the single excitation character and the deviation of the EOM-CCSD band gap 
from the experimental value. 
Figure \ref{fig:n1ip-vs-error-3d} depicts that we find small relative errors
with respect to experiment for $n_1^{\text{IP}}$ values above $\approx 95.5\%$, while all materials with
a smaller single excitation character (LiH, LiF and MgO) exhibit a sizable
disagreement compared to experiment.
Figure \ref{supp-fig:r1-vs-nk-ip} shows that the single excitation character
monotonically decreases with increasing simulation cell size for the IP. The same trend, even though with
a smaller magnitude, can be observed for the EA (see Figure \ref{supp-fig:r1-vs-nk-ea}).
In the case of LiH, whose band gap we
find to deviate by $0.82\,\text{eV}$ from experiment this decrease of $n_1^{\text{IP}}$ is the most extreme,
dropping by $2\%$.
For MgO, the single excitation character of the 
IP is also relatively low ($94.2\%$) and comparable to LiH ($93.6\%$). 
In the case of IP-EOM-CCSD one can clearly see that the magnitude of the
single excitation character differ visibly between the materials which are in good agreement 
with experiment and those which exhibit sizable disagreements in the band gap value.
This is strong evidence that the single excitation character is a useful quantitative marker to determine whether the EOM-CCSD
method yields accurate results for a given system.
%\section{Conclusion}\label{sec:conclusion}
\newline
\emph{Conclusion.} ---
%%%SHORTENED VERSION
We presented an efficient approach to extrapolate EOM-CCSD band gaps
to the TDL using the scaling of $G_0W_0$ gaps with respect to system size.
The precision of this approach was verified through agreement between 
FHI-aims and VASP calculations, while accuracy was assessed against experimental data and $GW^{\rm TC-TC}$ results. 
We found that the EOM-CCSD band gap quality correlates with the single excitation character of the quasi-particle excitation, 
with accuracy declining significantly for single excitation characters below 95\% - a trend confirmed in the alternating hydrogen chain.
This paves the way for predictions of band gaps with controlable accuracy.
Future work should explore the inclusion higher-order excitation processes in the $\hat{T}$ and $\hat{R}$ operators and
the utilization of non-HF-based single particle states to address these limitations.
\newline
\emph{Acknowledgment.} ---
%Support from the European Union's Horizon 2020 research and innovation program under Grant Agreement No. 951786 (The NOMAD CoE)
%is gratefully acknowledged.
This project was supported by TEC1p [the European Research Council (ERC) Horizon 2020 research
and innovation program, Grant Agreement No.740233] and Grant Agreement No. 101087184.
TS acknowledges support from the Austrian Science Fund (FWF) [DOI:10.55776/ESP335].
This work has partly been supported by the European Union
Horizon 2020 research and innovation program under the
Grant Agreement No 951786 (NOMAD CoE).
The computational results presented have partly been achieved using the Vienna Scientific Cluster (VSC)
and the Max Plank Computing and Data Facility (MPCDF).

\nocite{*}

\bibliography{main}

\makeatletter\@input{supplementaryxx.tex}\makeatother
\end{document}

%\maketitle
\widetext
\begin{center}
%\textbf{\large Supporting Information: Exploring the accuracy of the EOM-CC band gap in the thermodynamic limit}
\end{center}
\section{Structural parameters and band gap position}
\begin{table}[h]
	\centering
	\caption{Lattice parameter(s), space group and position of valence band maximum (VBM) and conduction band minimum (CBM)
	in reciprocal space in relative coordinates for the materials studied in this work.}
	\begin{tabular}{lllll}
		\toprule
		Material & Lattice parameter (\AA) & Space group & $\bm{k}_{\mathrm{VBM}}$ & $\bm{k}_{\mathrm{CBM}}$\\
		\midrule
		LiH & 4.084 & $Fm\bar{3}m$ & $(0.0, 0.5, 0.5)$ & $(0.0, 0.5, 0.5)$\\
		MgO & 4.207 & $Fm\bar{3}m$ & $(0.0, 0.0, 0.0)$ & $(0.0, 0.0, 0.0)$\\
		BP  & 4.538 & $F\bar{4}3m$ & $(0.0, 0.0, 0.0)$ & $(0.0, 0.41\bar{6}, 0.41\bar{6})$\\
		C   & 3.567 & $Fd\bar{3}m$ & $(0.0, 0.0, 0.0)$ & $(0.0, 0.\bar{3}, 0.\bar{3})$ \\
		BN  & 3.607 & $F\bar{4}3m$ & $(0.0, 0.0, 0.0)$ & $(0.5, 0.0, 0.5)$\\
		Si  & 5.430 & $Fd\bar{3}m$ & $(0.0, 0.0, 0.0)$ & $(0.41, 0.0, 0.41)$\\
		LiCl& 5.106 & $Fm\bar{3}m$ & $(0.0, 0.0, 0.0)$ & $(0.0, 0.0, 0.0)$\\
		LiF & 4.010 & $Fm\bar{3}m$ & $(0.0, 0.0, 0.0)$ & $(0.0, 0.0, 0.0)$\\
		\bottomrule
	\end{tabular}
	\label{tab:latt-params-and-shifts}
\end{table}

The lattice parameters were taken from Table 2 in Reference~\cite{gruneis2010second}. 
The $k$-shifts for the positions of the conduction band minimum (CBM) and the  valence band maximum (VBM)
were determined using the PBE-DFA.
In the case of diamond (C), the conduction band minimum determined via PBE was found to be
at $\approx (0.0, 0.3621, 0.3621)$. Due to limitations of the Hartree-Fock (HF) method in FHI-aims on performing
calculations with a shifted $k$-mesh, $k$-shifts in FHI-aims had to be
performed via down-sampling from a finer mesh. Since the required $k$-mesh, which would be necessary
to down-sample to that $k$-point would be impractically large, the conduction band minimum of C was
evaluated at $\bm{k}_{\mathrm{CBM}} = (0.0, 0.\bar{3}, 0.\bar{3})$ as specified in Table 
\ref{tab:latt-params-and-shifts}. The resulting deviation of the band gap was quantified using the PBE-DFA
and HF theory and corresponds to $20\,\mathrm{meV}$ and $52\,\mathrm{meV}$, respectively.
For LiCl and LiF, the valence band maxima predicted by PBE are slightly shifted away from the $\Gamma$-point, $\bm{k}_{\mathrm{VBM}} = (0.07, 0.07, 0.14)$ for LiCl and $\bm{k}_{\mathrm{VBM}} = (0.06, 0.06, 0.11)$ for LiF. Due to the small magnitude of that shift and the very flat dispersion of the valence band around the $\Gamma$-point, the EOM-CCSD calculations were performed
using the $\Gamma$-point as the valence band maximum. On the PBE level of theory this 
approximation yields a deviation of less than $10\,\text{meV}$.
All EOM-CCSD calculations were performed using super cells, for which the $k$-shifts in Table \ref{tab:latt-params-and-shifts} were scaled appropriately. 

\section{Computational details}
\subsection{FHI-aims}
All $G_0W_0$@HF and EOM-CCSD calculations to determine the bulk-limit of the EOM-CCSD band gaps
were performed using the loc-NAO-VCC-2Z~\cite{zhang2019main} basis set to which a $f$-, $g$- and $h$-type auxiliary basis
function of effective charge $1.0$ was added manually to ensure sufficient completeness 
of the auxiliary basis. To perform the CC calculations, the converged HF quantities from FHI-aims
were subsequently post-processed by the CC-aims interface~\cite{moerman2022interface}, which generated the necessary input for the 
Cc4s software package~\cite{cc4s}.

To compute the EOM-CCSD band gaps the HF single-particle wave functions were obtained by 
down-sampling from a $k$-mesh of size $10\times 10\times 10$ or more. 
This was done because the current implementation
of FHI-aims does not allow to perform shifts of the $k$-mesh in combination with the HF method,
which, however, is necessary to sample the conduction band minimum of two of the investigated systems
(diamond and boron phosphide).
By performing the HF calculation on a denser $k$-grid and subsequently downsampling the CC
calculation to a coarser (and in these cases shifted) grid, this limitation can be circumvented.
%Since the VASP code is not limited by such a constraint, no down-sampling was employed for the
%calculations involving VASP.

The EOM-CCSD finite-size convergence study for LiH involved isotropic supercells of size 
$2\times2\times2-5\times5\times5$, while for the
other materials $2\times2\times2-4\times4\times4$ supercells were employed. While the EOM-CCSD
band gap was computed at $\bm{k}_{\mathrm{VBM}}$ and $\bm{k}_{\mathrm{CBM}}$ 
from Table \ref{tab:latt-params-and-shifts}, due to limitations of the $G_0W_0$@HF implementation
of FHI-aims, all $G_0W_0$ calculations related to the study of the 
band gap finite-size convergence were performed for the direct $\Gamma\to\Gamma$ band gap.

The BSIE was estimated by performing $2\times2\times2$ EOM-CCSD super cell calculations
at the (appropriately scaled) $\bm{k}_{\mathrm{VBM}}$ and $\bm{k}_{\mathrm{CBM}}$ 
from Table \ref{tab:latt-params-and-shifts} using the loc-NAO-VCC-3Z and -4Z basis sets. 
For all studied materials, the change of the EOM-CCSD band gap between the 3Z and 4Z was found to be 
less than $130\,\mathrm{meV}$ (see Table \ref{tab:bsie-summary}). Similarly to the BSIE correction performed in the EOM-CCSD
study by Vo \emph{et al.}~\cite{vo2024performance}, the BSIE in this work was estimated as the difference
between the 2Z basis result used in the study of the finite-size convergence and the 4Z result for the
$2\times2\times2$ super cell and was added to the 2Z-based bulk-limit result of the EOM-CCSD band gap.

To remain consistent with the treatment of the long-range contribution of Coulomb potential
in the underlying HF calculations all $G_0W_0$ and EOM-CCSD calculations were performed using
the cut-Coulomb potential~\cite{levchenko2015hybrid}. 

\subsection{VASP}
The PAW-based EOM-CCSD calculations involved in the study of the finite-size convergence
of the band gap were performed with a basis set of 6 virtual orbitals per 
occupied orbital ($N_v/N_o=6$) using super cells of size $2\times2\times2 - 5\times5\times5$
for LiH and with $N_v/N_o=3$ for the $2\times2\times2$ and $3\times3\times3$ super cells
for MgO, BP and C. In contrast to FHI-aims, VASP does not have a restriction
on performing HF calculations for shifted $k$-meshes, so that no down-sampling was employed.
Similar to FHI-aims, the $G_0W_0$ implementation in VASP is not compatible with shifted
$k$-meshes, so that all $GW$ calculations were performed for the $\Gamma\to\Gamma$ band gap.
The long-range contributions to the Coulomb potential for both EOM-CCSD and the $GW$ calculations
were approximated using the probe-charge Ewald method~\cite{massidda1993hartree}.

In analogy to the approach pursued for FHI-aims, the BSIE of the PAW-based
calculations was quantified by computing the EOM-CCSD band gap of the $2\times2\times2$
super cell using the aforementioned basis set of size $N_v/N_o=6$ (LiH) or $N_v/N_o=3$ (MgO, BP, C) and a much bigger basis set
of $N_v/N_o=19$, serving as an estimate to the complete basis limit. The BSIE estimate was taken
as the difference between the band gap values of these two basis set sizes and is tabulated
in Table \ref{tab:bsie-summary}. That difference was added to the finite-size extrapolated EOM-CCSD band gap
to obtain the final gap value.

The $G_0W_0$@HF calculations were performed employing the single step $GW$ procedure 
in VASP (available in versions $>$ 6.3), which automatically determines all necessary 
computational parameters.

Both the $G_0W_0$ and EOM-CCSD calculations were performed using the 
\texttt{Li\_GW}, \texttt{H\_GW}, \texttt{Mg\_GW}, \texttt{O\_GW\_new}, 
\texttt{P\_GW}, \texttt{B\_GW} and \texttt{C\_GW} POTCARs and the maximal
recommended kinetic energy cut-off (\texttt{ENMAX}) therein. 

%TODO: Add GW-TC-TC details
The self-consistent vertex-corrected $GW$ calculations
were performed using the $GW^{\mathrm{TC-TC}}$ implementation
in VASP~\cite{shishkin2007accurate}. In contrast to the
previously discussed $G_0W_0$@HF calculations, the 
finite-size convergence of the $GW^{\mathrm{TC-TC}}$
calculations was accelerated using head- and wing-corrections.
$GW^{\mathrm{TC-TC}}$ for up to $6\times6\times6$ $k$-points
were performed and extrapolated to the complete basis set limit.
The final band gap values are estimated to have a remaining 
uncertainty of $50\,\text{meV}$.

The zero-point renormalizations (ZPR) were taken directly from Reference
\cite{engel2022zero} while the values for LiH, LiCl and BP were computed following the same perturbative approach.
It should be pointed out that in Ref.~\cite{engel2022zero} a lattice parameter of 4.211 \AA{} was used for MgO, 3.536 \AA{} for C and 4.055 \AA{} for LiF.
%For the EOM-CC electronic calculations 4.207 \AA for MgO and 3.567 \AA for C were used.
Also, the LDA exchange correlation functional was used for the C calculation and PBE for MgO.

For the LiH, LiCl and BP calculations we used a PBE exchange correlation functional, cutoff energy of 520 eV and
supercells made of 3x3x3 copies of the conventional cell
and 2x2x2 $k$-point sampling for BP and LiCl and 3x3x3 $k$-point sampling for LiH to compute the electron-phonon potential.
We included 474 bands for LiH, 934 bands for LiCl and 640 bands for BP in the sum-over-states for the ZPR calculation and a small
imaginary complex shift of 0.01 eV.
A summary of the parameters and results of the ZPR calculations is shown in Table \ref{tab:params-zpr}.

\begin{table}[h]
	\centering
	\caption{Parameters for the ZPR calculation of BP, LiH and LiCl.}
	\begin{tabular}{llllllll}
		\toprule
		Material & Lattice parameter (\AA) & $\bm{k}_{\mathrm{VBM}}$ & $\bm{k}_{\mathrm{CBM}}$ & $\epsilon_{xx}^\infty$ & $Z_{xx}^*$ & ZPR (meV) & POTCAR\\
		\midrule
		BP  & 4.538 & $(0.0, 0.0, 0.0)$ & $(0.0, 0.4149, 0.4149)$ & 9.150 & 0.508 & 106 & \texttt{B\_GW\_new} \texttt{P\_GW}\\
		BP  & 4.538 & $(0.0, 0.0, 0.0)$ & $(0.0, 0.4149, 0.4149)$ & 9.158 & 0.514 & 105 & \texttt{B\_GW} \texttt{P\_GW}      \\
		LiH & 4.084 & $(0.0, 0.5, 0.5)$ & $(0.0, 0.5, 0.5)$       & 4.294 & 1.027 & 444 & \texttt{Li\_sv\_GW} \texttt{H\_GW} \\
		LiH & 4.084 & $(0.0, 0.5, 0.5)$ & $(0.0, 0.5, 0.5)$       & 4.249 & 1.027 & 450 & \texttt{Li\_GW} \texttt{H\_GW}     \\
		LiCl& 5.106 & $(0.0, 0.0, 0.0)$ & $(0.0, 0.0, 0.0)$       & 2.979 & 1.184 & 543 & \texttt{Li\_sv\_GW} \texttt{Cl\_GW}   \\
		\bottomrule
	\end{tabular}
	\label{tab:params-zpr}
\end{table}

\subsection{NWChem}
The calculations of the finite hydrogen chain were performed 
with a def2-SVP basis set using the NWChem code. Both the IP/EA-EOM-CCSD
and the $\Delta$CCSD(T) calculations used a spin-restricted HF starting-point,
which in the case of the charged $N-1$ and $N+1$ systems was
given by the spin-restricted open-shell HF (ROHF) wave function and eigenvalues.

\section{Estimation of the basis set incompleteness error}\label{sec:bsie}

\begin{table}[ht]
        \centering
        \caption{Change of $2\times2\times2$ EOM-CCSD band gap with increasing basis set
	for FHI-aims and VASP. For FHI-aims, the change of the loc-NAO-VCC-2Z and -3Z based 
	band gap relative to the loc-NAO-VCC-4Z basis set is shown. 
	For VASP the relative change of the EOM-CCSD band gap obtained with $N_v/N_o=6$ (LiH)
	or $N_v/N_o=3$ (MgO, BP, C) compared to a basis set of size $N_v/N_o=19$ is shown.
	All entries are in eV.}
	\begin{tabular}{p{3cm}cccl}
                \toprule
                \multicolumn{1}{ c }{Material}&
		\multicolumn{2}{ c }{BSIE wrt. 4Z (FHI-aims)}&
		\multicolumn{1}{ c }{BSIE wrt. $N_v/N_0=19$ (VASP)}\\ 
		\cmidrule(lr){2-3} \cmidrule{4-4}
                    &
                2Z &
                3Z &
                $N_v/N_o=3$ \\
                \midrule
                LiH & -0.029  & 0.005 & 0.089 \\
                MgO & -0.446  & 0.018 & 0.139 \\
                BP  & -0.331  & -0.057 & 0.125 \\
                C   & -0.556  & -0.124 & 0.102 \\
		BN  & ---     & ---    & 0.195 \\
		Si  & ---     & ---    & 0.170 \\
		LiCl& ---     & ---    & 0.052 \\
		LiF & ---     & ---    & -0.003 \\ 
                \bottomrule
        \end{tabular}
	
	\label{tab:bsie-summary}
\end{table}

\newpage
\section{$G_0W_0$@HF and EOM-CCSD convergence using the PAW method}

\begin{figure}[ht]
\centering
\begin{center}
    \includegraphics[width=.4\linewidth]{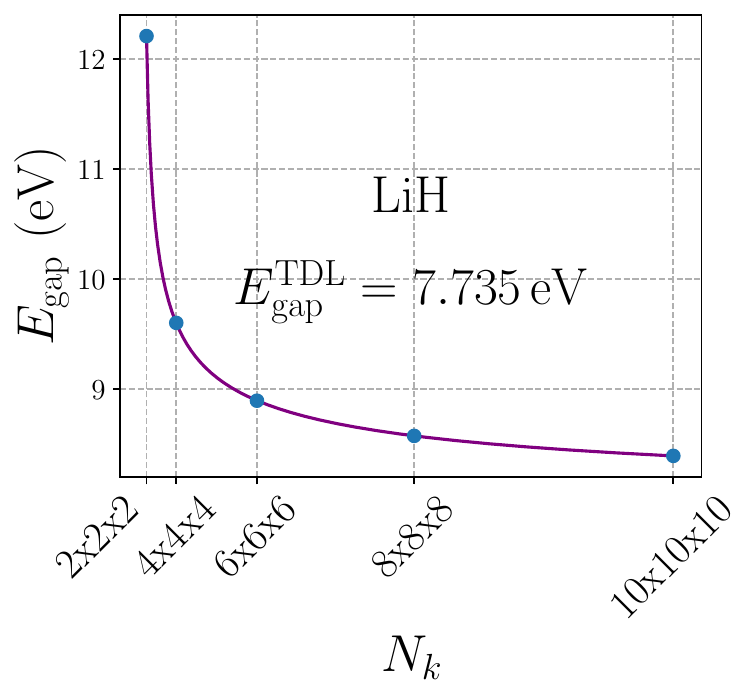}%
    \includegraphics[width=.4\linewidth]{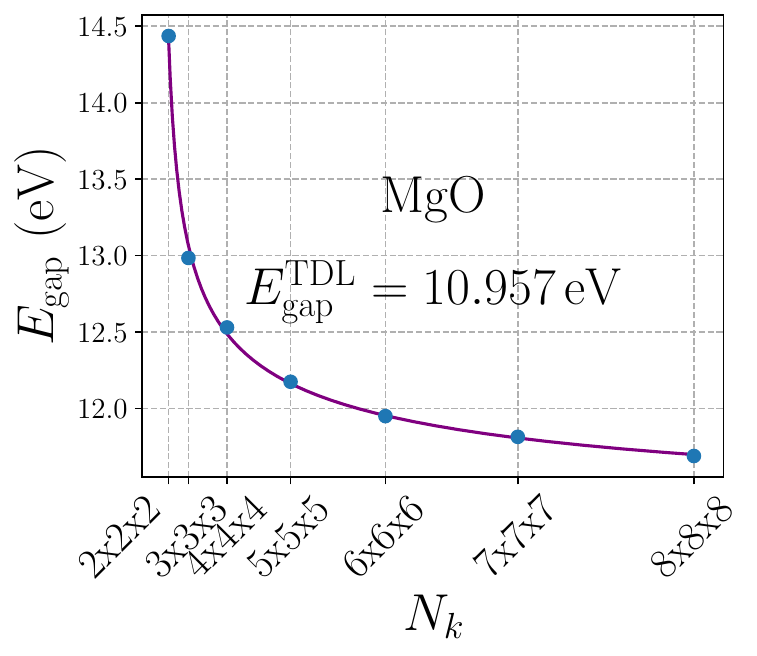}
    \includegraphics[width=.4\linewidth]{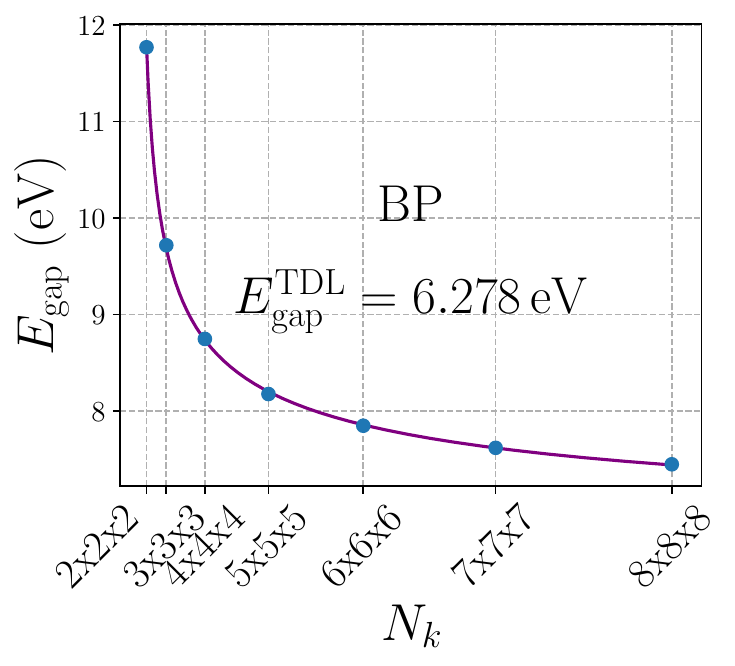}%
    \includegraphics[width=.4\linewidth]{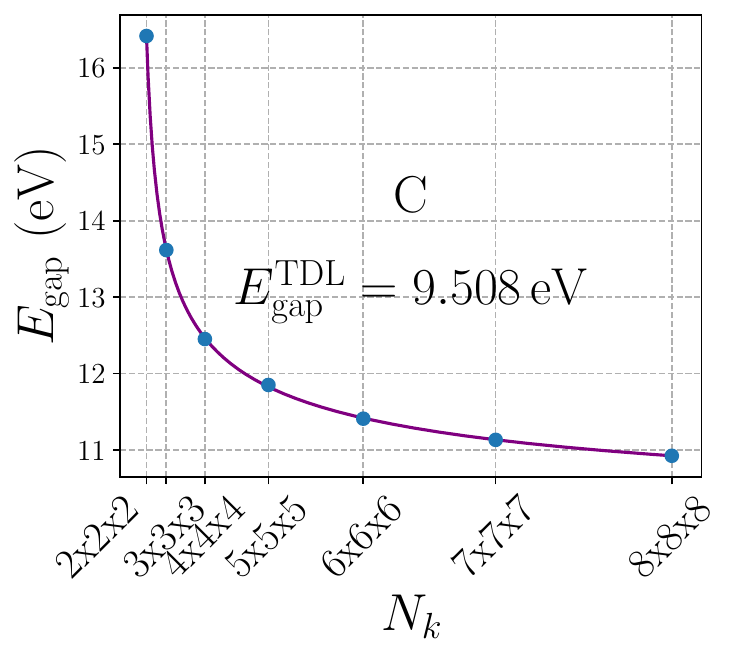}
	\phantomcaption
\end{center}
\end{figure}

\begin{figure}[ht]
    \ContinuedFloat
    \centering
   \begin{center}
    \includegraphics[width=.4\linewidth]{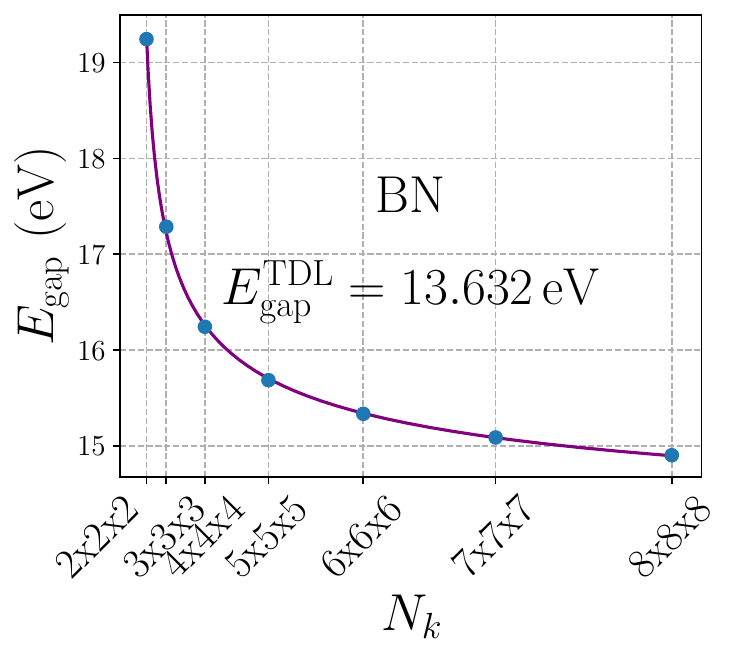}
    \includegraphics[width=.4\linewidth]{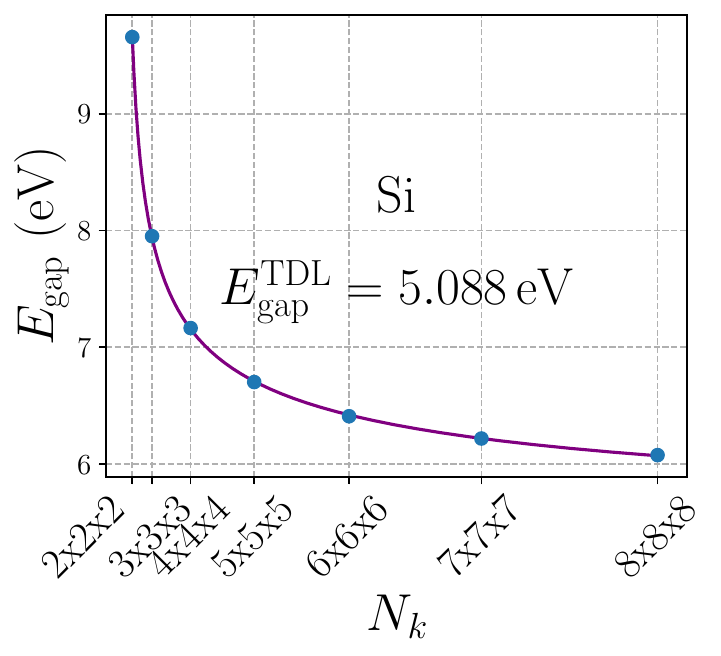}
    \includegraphics[width=.4\linewidth]{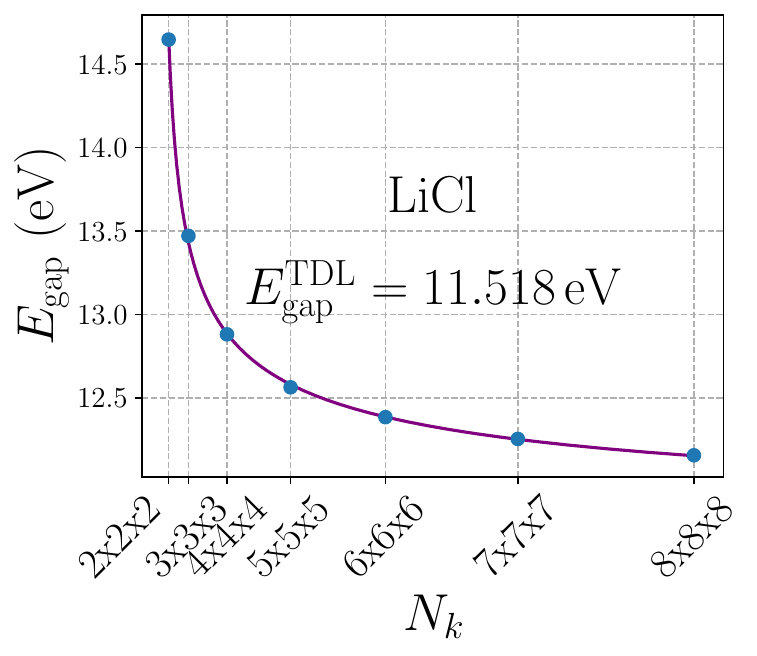}
    \includegraphics[width=.4\linewidth]{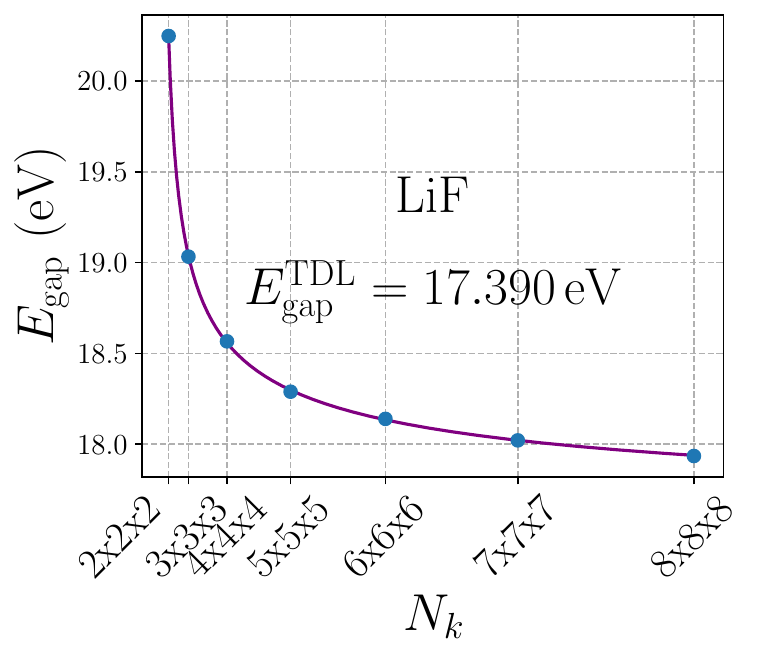} 
   \end{center}
    \caption{Convergence of the VASP $G_0W_0$@HF band gaps with respect to the $\bm{k}$-mesh. The data points were fitted and
    extrapolated using the expression
    $E_{\mathrm{gap}}(N_k) = E_{\mathrm{gap}}^{\mathrm{TDL}} + AN_k^{-1/3} + BN_k^{-2/3} + CN_k^{-1}$.}
    \label{fig:3d-gw-extrapolations-vasp}
\end{figure}

\begin{figure}[ht]
    \centering
   \begin{center}
    \includegraphics[width=.4\linewidth]{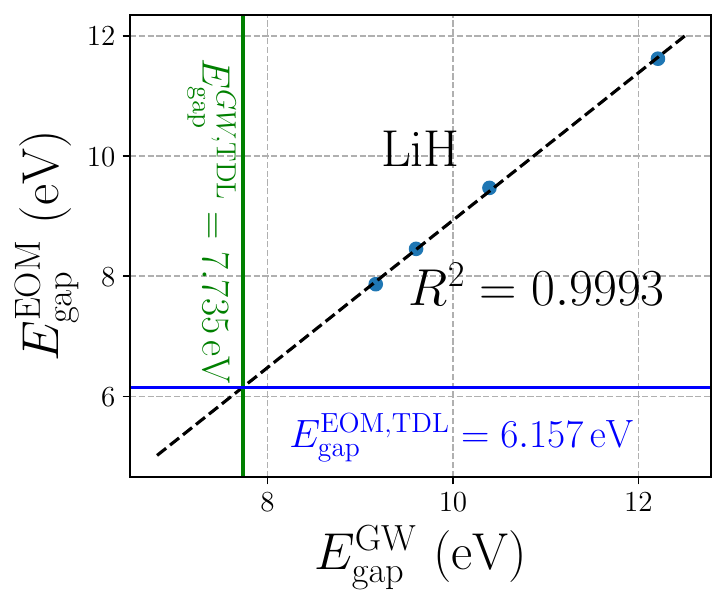}%
    \includegraphics[width=.4\linewidth]{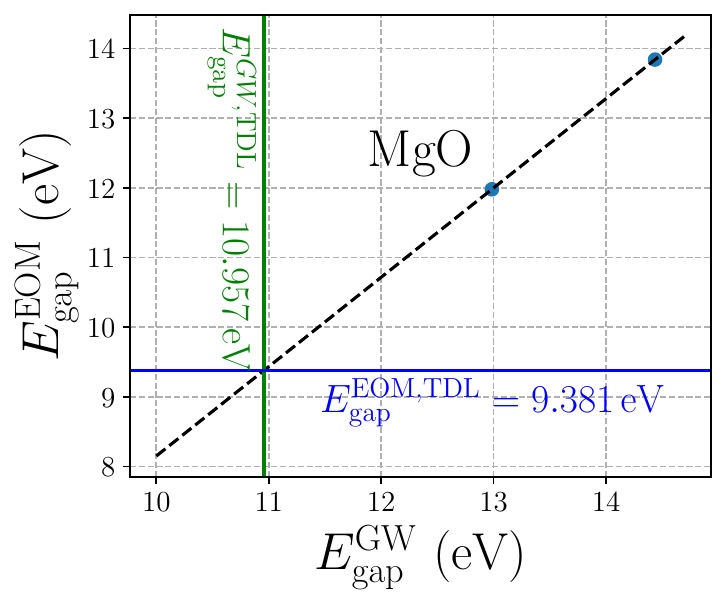}
    \includegraphics[width=.4\linewidth]{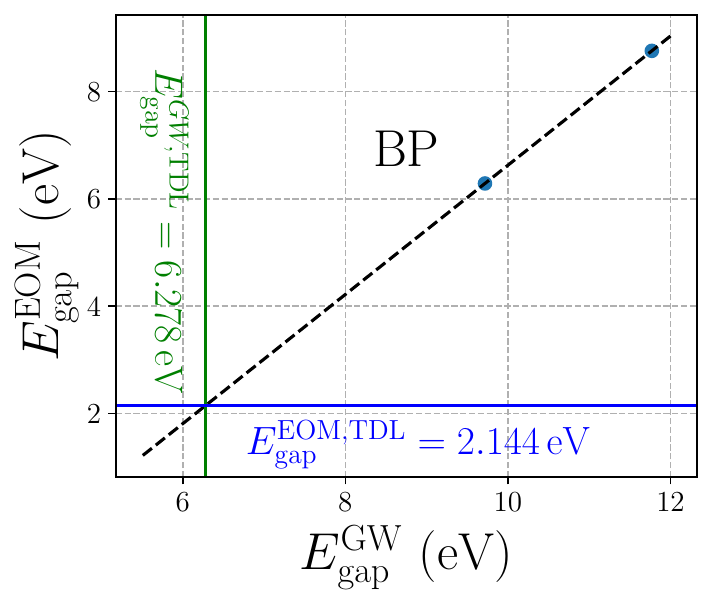}%
    \includegraphics[width=.4\linewidth]{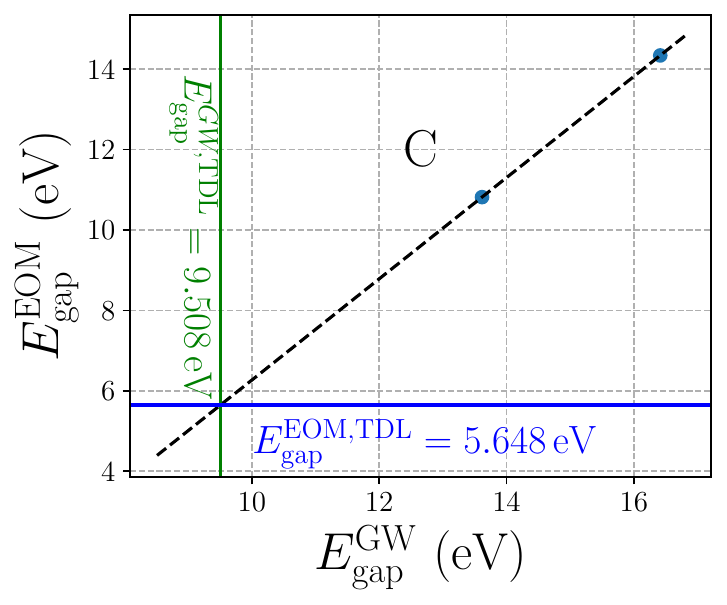}
    \includegraphics[width=.4\linewidth]{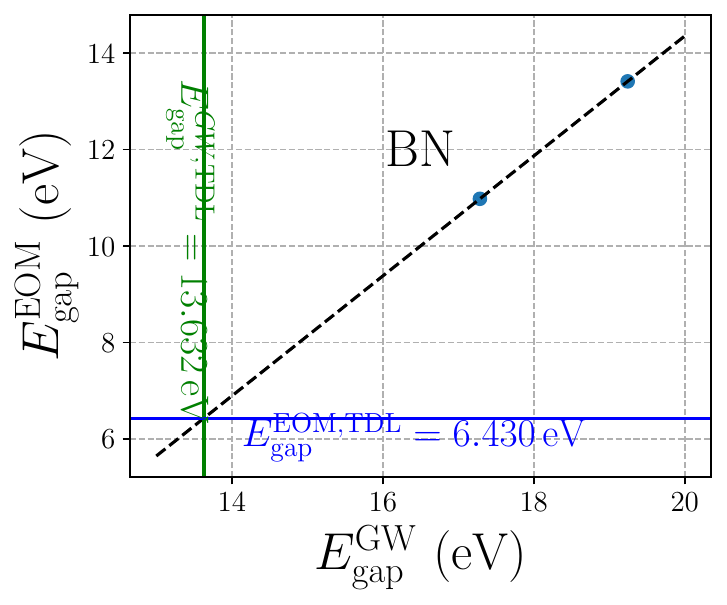}
    \includegraphics[width=.4\linewidth]{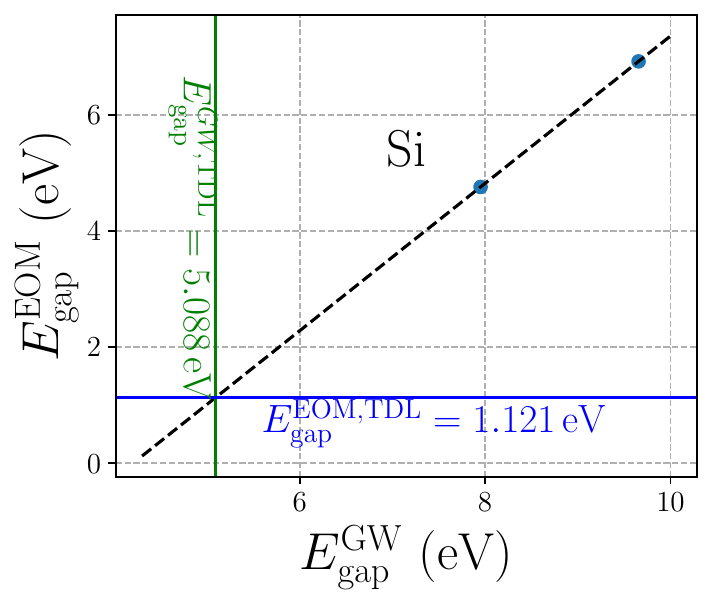}
	   \phantomcaption
   \end{center}
\end{figure}

\begin{figure}[ht]
    \ContinuedFloat
    \centering
   \begin{center}
    \includegraphics[width=.4\linewidth]{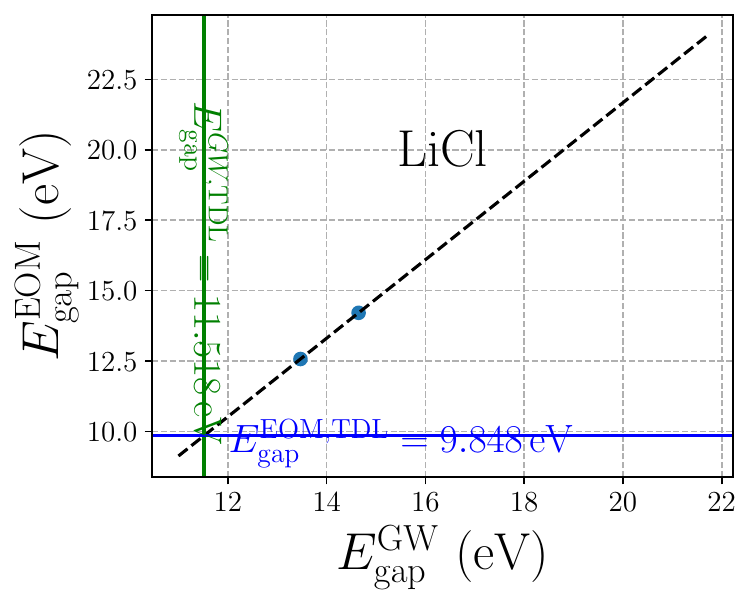}
    \includegraphics[width=.4\linewidth]{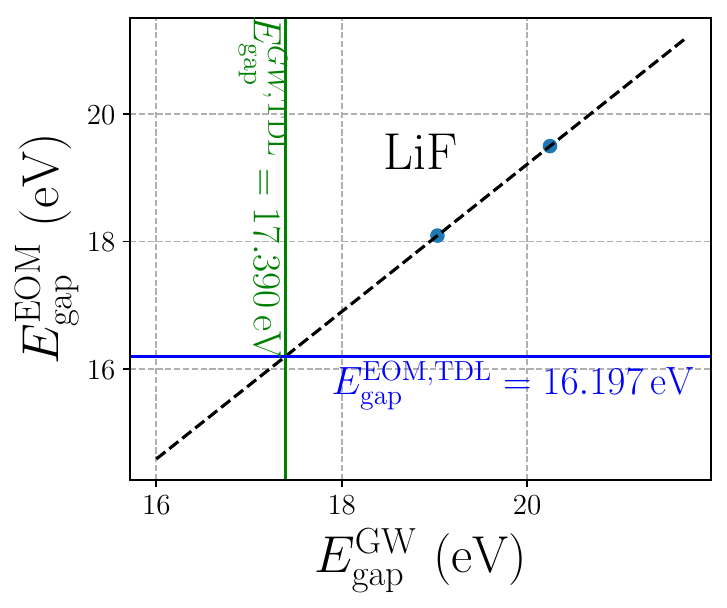}  
   \end{center}
    \caption{Correlation of the EOM-CCSD and $G_0W_0$@HF band gaps
    convergence
    using VASP with increasing supercell size (EOM-CCSD) and $\bm{k}$-mesh density ($G_0W_0$@HF). For LiH
    isotropic supercells of size $2\times2\times2-5\times5\times5$ were used, while for the other materials $2\times2\times2$ and $3\times3\times3$ supercells were employed. For all four
    systems the band gap converges from above, so that system size increases
    from right to left. At the
    TDL value of the $GW$
    band gap, obtained in the previous section via extrapolation, a vertical
    line was drawn. The resulting
    TDL estimate of the
    EOM-CCSD band gap is marked
    by a horizontal line.}
    \label{fig:3d-eom-gw-correlation-vasp}
\end{figure}

\begin{figure}
	\centering
	\subfloat[][]{\includegraphics[width=0.4\textwidth]{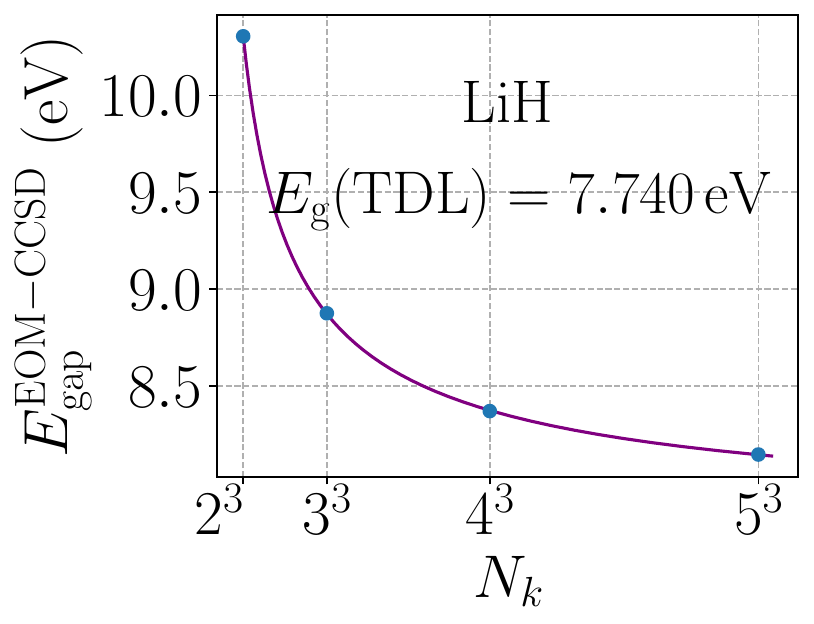}\label{fig:pure-eom-LiH-all-orders-aims}}
	\subfloat[][]{\includegraphics[width=0.4\textwidth]{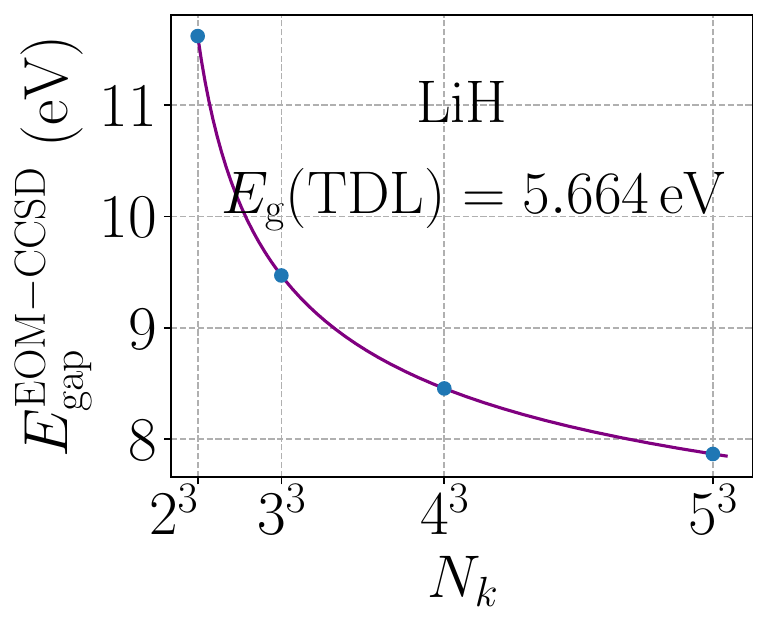}\label{fig:pure-eom-LiH-all-orders-vasp}}
	\caption{Convergence of the EOM-CCSD band gap of LiH and extrapolation via $AN_k^{-1/3} + BN_k^{-2/3} + CN_k^{-1}$ based on results from a) FHI-aims and b) VASP}
	\label{fig:pure-eom-LiH-all-orders}
\end{figure}

\begin{figure}
	\centering
	\subfloat[][]{\includegraphics[width=0.4\textwidth]{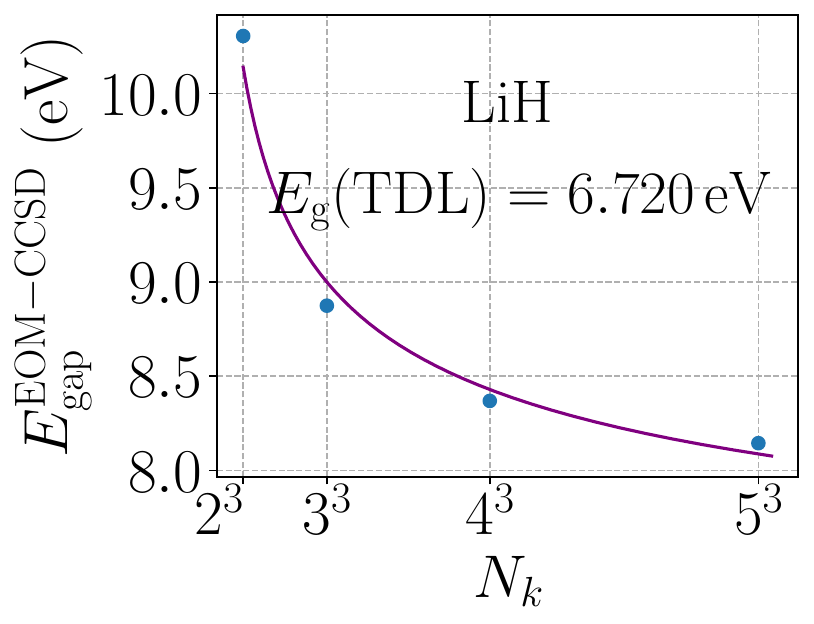}\label{fig:pure-eom-LiH-lead-orders-aims}}
	\subfloat[][]{\includegraphics[width=0.4\textwidth]{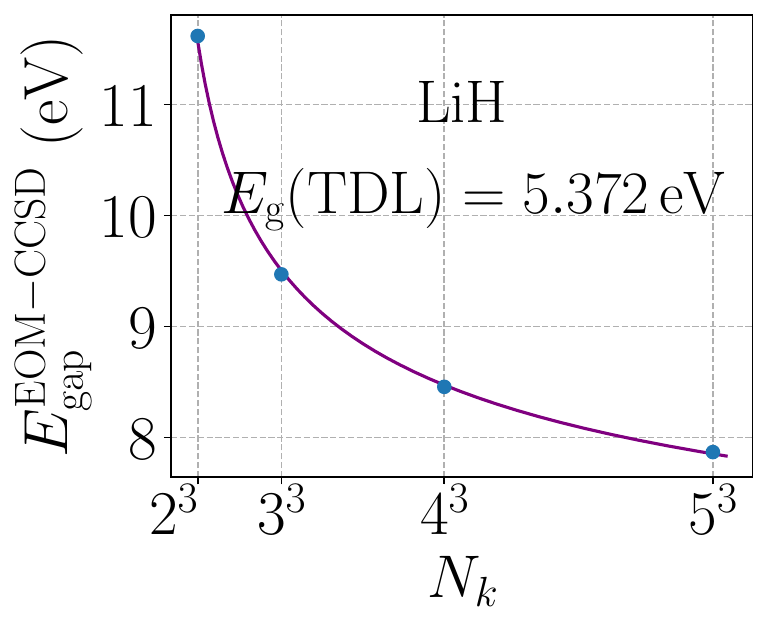}\label{fig:pure-eom-LiH-lead-orders-vasp}}
	\caption{Convergence of the EOM-CCSD band gap of LiH and extrapolation only via the leading order term $AN_k^{-1/3}$ based on results from a) FHI-aims and b) VASP}
	\label{fig:pure-eom-LiH-lead-orders}
\end{figure}

\begin{figure}
	\centering
	\subfloat[][]{\includegraphics[width=0.4\textwidth]{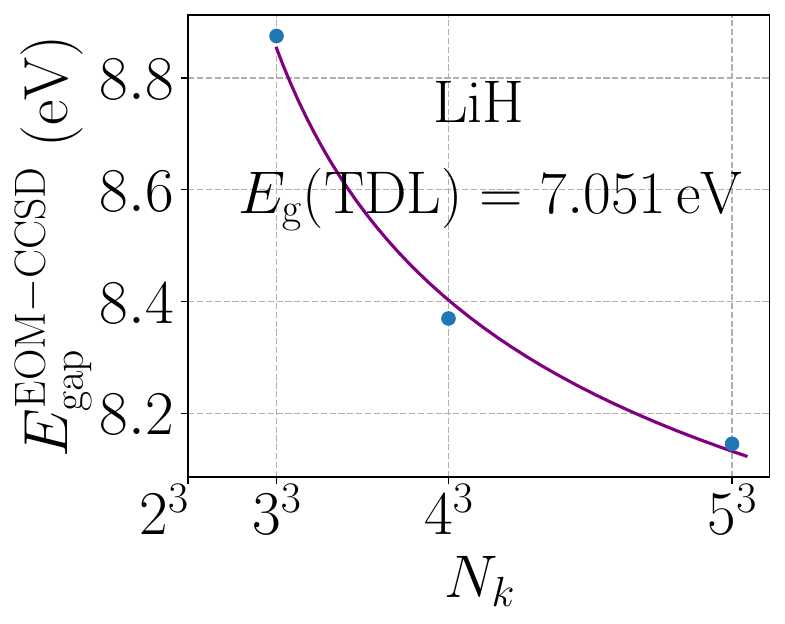}\label{fig:pure-eom-LiH-lead-orders-aims-3-5}}
	\subfloat[][]{\includegraphics[width=0.4\textwidth]{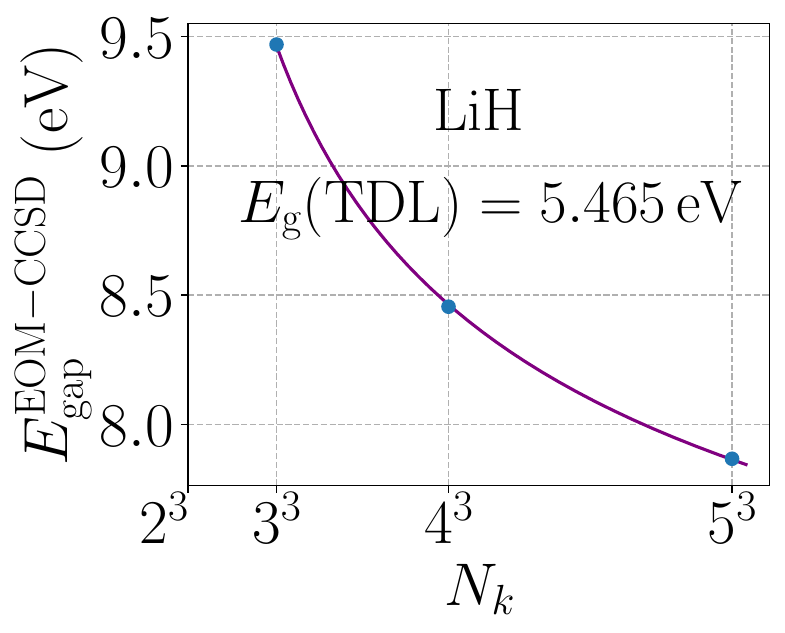}\label{fig:pure-eom-LiH-lead-orders-vasp-3-5}}
	\caption{Convergence of the EOM-CCSD band gap of LiH and extrapolation only via the leading order term $AN_k^{-1/3}$ based on $3\times3\times3$ -- $5\times5\times5$ results from a) FHI-aims and b) VASP}
	\label{fig:pure-eom-LiH-lead-orders-3-5}
\end{figure}

\newpage
\section{Single excitation character of EA-EOM-CCSD}

\begin{figure}[ht]
	\includegraphics[]{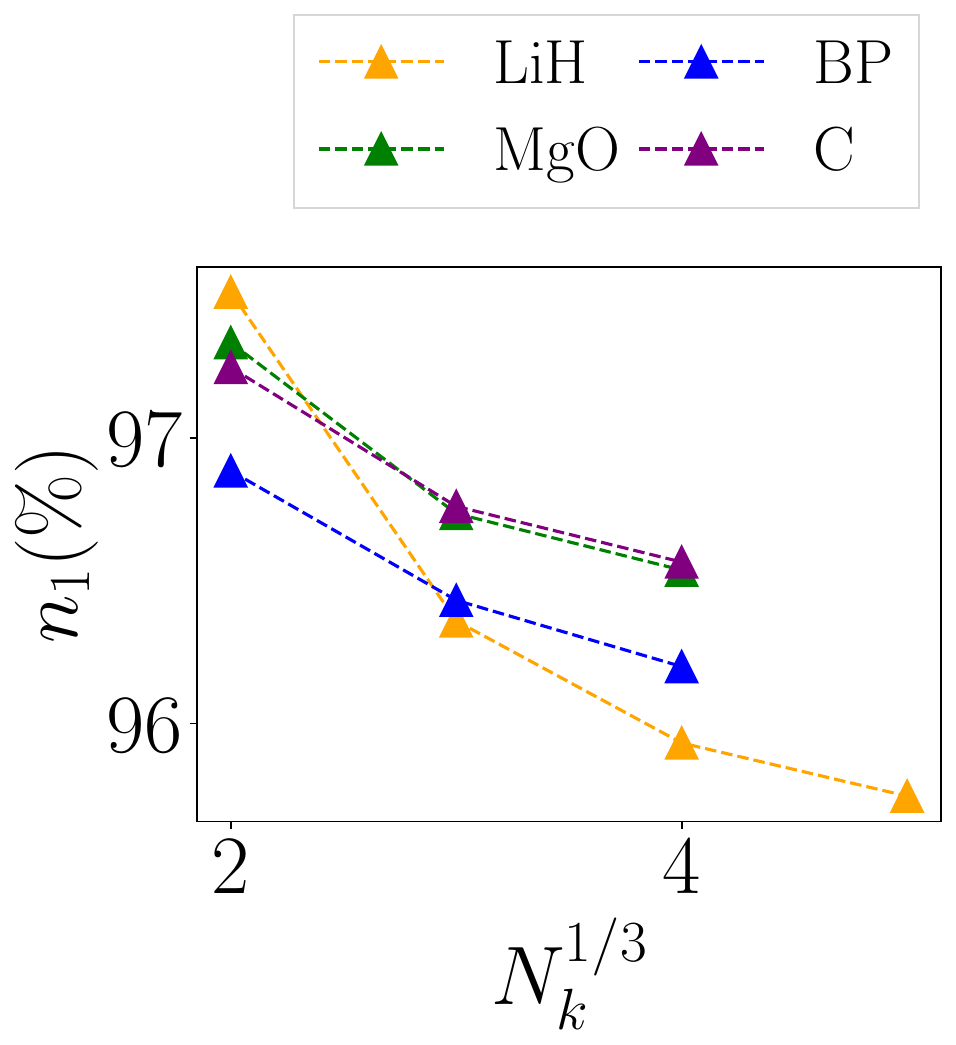}
	\caption{Change of single excitation character of the EA quasi-particle for the
	3D bulk materials with respect to the super cell size.}
	\label{fig:r1-vs-nk-ea}
\end{figure}

\begin{figure}[ht]
	\includegraphics[]{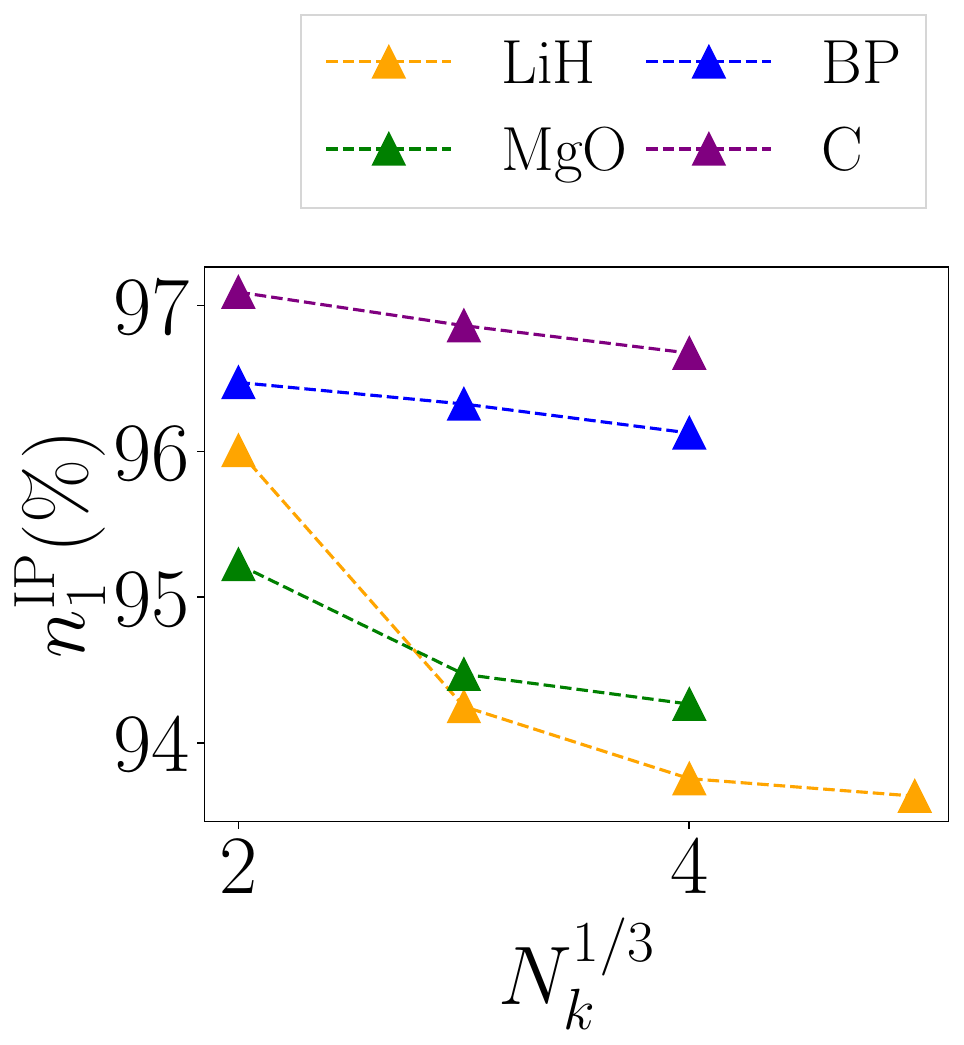}
	\caption{Change of single excitation character of the IP quasi-particle for the
	3D bulk materials with respect to the super cell size.}
	\label{fig:r1-vs-nk-ip}
\end{figure}

\newpage
\section{IP- and EA-EOM-CCSD benchmark for small molecules}
\begin{table}[t]
    \centering
    \begin{tabular}{llllll}
    \toprule
        system & $n$& loc-NAO-VCC-$n$Z & NAO-VCC-$n$Z & cc-pV$n$Z~\cite{musial2003equationip} & aug-cc-pV$n$Z~\cite{musial2003equationip}\\
        \midrule
        N\textsubscript{2} & 2 & 15.49 & 15.52 & 15.18 & 15.43\\
                           & 3 & 15.73 & 15.64 & 15.56 & 15.65\\
                           & 4 & 15.69 & 15.70 & 15.68 & 15.71\\
        \midrule
                           &CBS&       & 15.75 & 15.78 & 15.80\\
        \midrule
        \midrule
        F\textsubscript{2} & 2 & 16.10 & 15.58 & 15.10 & 15.40\\
                           & 3 & 15.86 & 15.66 & 15.49 & 15.61\\
                           & 4 & 15.81 & 15.72 & 15.66 & 15.72\\
        \midrule
                           &CBS&       & 15.76 & 15.80 & 15.82\\
        \midrule
        \midrule
        CO & 2 & 14.23 & 14.09& 13.81 & 13.99\\
           & 3 & 14.29 & 14.16& 14.13 & 14.18\\
           & 4 & 14.24 & 14.22& 14.22 & 14.23\\
    \midrule
           &CBS&       & 14.27& 14.30 & 14.30\\
    \bottomrule
    \bottomrule
    \end{tabular}
    \caption{Ionization potentials for N\textsubscript{2}, F\textsubscript{2}
    and CO using NAOs and GTOs. The molecular geometries
    and the the GTO-based values of the ionization potentials were
    taken from Reference~\cite{musial2003equationip}. The CBS extrapolation
    of the NAO-based results was performed employing
    a cubic two-point extrapolation expression using the NAO-VCC-3Z
    and NAO-VCC-4Z result.
    The results are given in eV.}
    \label{tab:molecules-ip}
\end{table}

\begin{table}[t]
    \centering
    \begin{tabular}{llllll}
        \toprule
        system & $n$& loc-NAO-VCC-$n$Z & NAO-VCC-$n$Z & cc-pV$n$Z~\cite{musial2003equationea} & aug-cc-pV$n$Z~\cite{musial2003equationea}\\
        \midrule
        C\textsubscript{2} & 2 & 3.10 & 3.08 & 2.53 & 3.13\\
                           & 3 & 3.19 & 3.19 & 3.07 & 3.30\\
                           & 4 & 3.29 & 3.26 & 3.24 & 3.35\\
        \midrule
                           &CBS&       & 3.32 & 3.34 & 3.38\\
        \midrule
        \midrule
        O\textsubscript{3} & 2 & 1.63 & 1.43 & 0.61 & 1.57\\
                           & 3 & 1.48 & 1.68 & 1.31 & 1.77\\
                           & 4 & 1.73 & 1.76 & 1.63 & 1.85\\
        \midrule
                           &CBS&       & 1.82 & 1.82 & 1.95\\
        \bottomrule
        \bottomrule
    \end{tabular}
    \caption{Electron affinities for C\textsubscript{2} and O\textsubscript{3}
     using NAOs and GTOs. The molecular geometries
    and the the GTO-based values of the electron affinities were 
    taken from Reference~\cite{musial2003equationea}. The CBS extrapolation
    of the NAO-based results was performed employing
    a cubic two-point extrapolation expression using the NAO-VCC-3Z
    and NAO-VCC-4Z result.
    The results are given in eV.}
    \label{tab:molecules-ea}
\end{table}

%\nocite{*}
\clearpage
\bibliography{supplementary}